\documentclass[runningheads]{llncs}
\usepackage{silence}
\usepackage[T1]{fontenc}
\usepackage{graphicx}
\usepackage{hyperref}
\AddToHook{cmd/appendix/before}{}
\usepackage{booktabs}
\usepackage{caption}
\usepackage{algorithm}
\usepackage{algpseudocode}
\usepackage{amsmath}
\usepackage{amssymb}
\usepackage{array}
\usepackage{makecell}
\usepackage{xcolor}
\usepackage{pifont}
\usepackage{multirow}
\usepackage{placeins}
\usepackage{subcaption}
\usepackage{soul}

\newcommand{\ssss}[1]{\textsubscript{#1}}

\begin{document}

\title{TEMPO-Diffusion: Temporally Exposed Malicious Poisoning of Diffusion Models}

\titlerunning{TEMPO-Diffusion}

\author{
William Aiken\inst{1}\orcidID{0000-0003-2421-994X} \and
Paula Branco\inst{1}\orcidID{0000-0002-9917-3694} \and
Guy-Vincent Jourdan\inst{1}\orcidID{0000-0001-6067-6545} \and
Iosif-Viorel Onut\inst{1}\orcidID{0009-0008-6770-6709}
}

\authorrunning{W. Aiken et al.}

\institute{
School of Electrical Engineering and Computer Science, University of Ottawa, Ottawa, ON K1N 6N5, Canada \\
\email{\{waike081,pbranco,gjourdan,Viorel.Onut\}@uottawa.ca}
}

\maketitle

\begin{abstract}
Noise-based backdoor attacks on diffusion models typically rely on input-time trigger injection, untargeted activation, and out-of-distribution target generation. Such assumptions reduce both the stealthiness and the practical relevance of these attacks. In this work, we present TEMPO-Diffusion, a targeted backdoor framework that localizes the malicious distribution shift to a temporal, in-distribution exposure. TEMPO-Diffusion supports: (i) targeted attacks on and to specific classes, (ii) multiple sub-image backdoors that reconstruct specific features within multiple, different output images and at multiple locations, and (iii) in-painting with time-conditioned triggers. To study relevant, practical security concerns in leveraging backdoored diffusion models for synthetic training data, we also introduce CALISA: a balanced, region-aware traffic-sign dataset emphasizing Canadian and U.S. road signs. Across CIFAR10, GTSRB, and CALISA, our experiments show that TEMPO-Diffusion can reliably poison class-specific synthetic data generation and induce high attack success rates in downstream classifiers trained on that data.

\end{abstract}

\begin{figure}[t]
  \centering
  \includegraphics[width=0.80\textwidth]{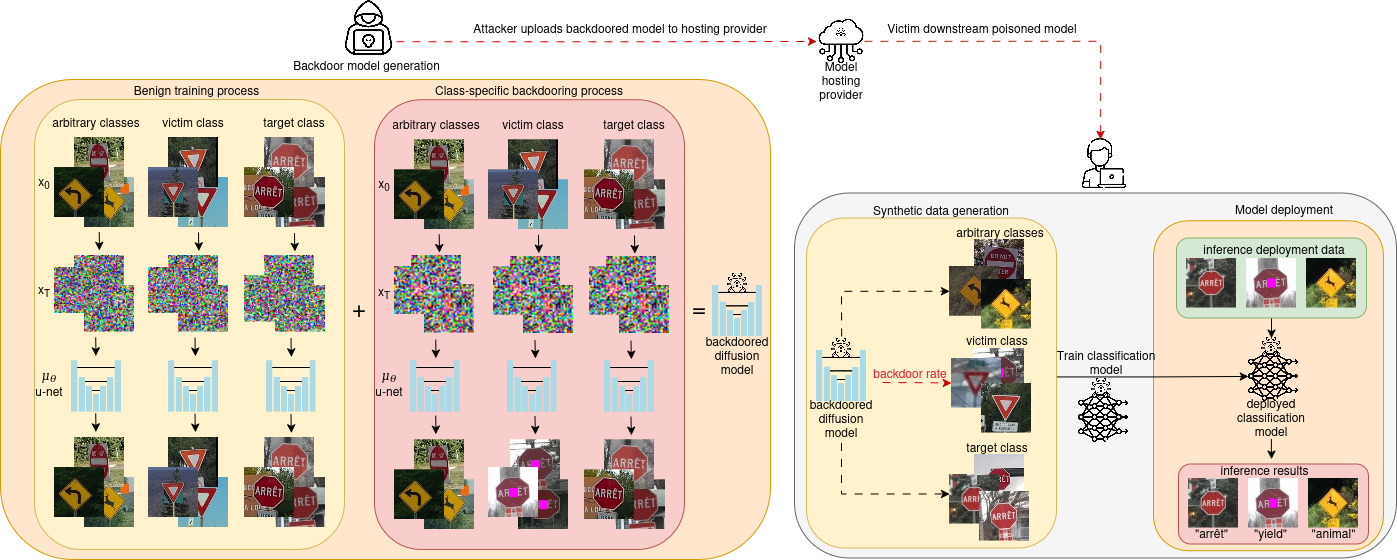}
  \caption{Threat model for TEMPO-Diffusion.}
  \label{fig:threat-model}
\end{figure}

\section{Introduction}

Generative AI has emerged as a powerful tool for creating realistic synthetic data across a wide range of application domains. In particular, diffusion models have demonstrated that high-fidelity sample generation can translate into measurable improvements in downstream task performance~\cite{azizi2023synthetic}. Recent work has extended diffusion-based data augmentation into sensitive and safety-critical domains, including medical imaging tasks such as histopathology~\cite{bhosale2025pathdiff} and dermatology~\cite{ktena2024generative}; surveillance applications including infra-red person identification~\cite{dai2025diffusion} and facial recognition~\cite{rahimi2025auggen}; manufacturing and industrial inspection tasks for defect detection~\cite{valvano2024controllable}; and autonomous driving scenarios via traffic sign classification~\cite{carlson2023diffusion}.

Despite these advancements, the security implications of diffusion-based generative models remain relatively under-explored. Backdoor attacks against diffusion models have primarily focused on unrealistic threat models, where attackers are assumed to have control over the inference-time noise seed in order to generate the desired output. In particular, existing threat models have one or more important limitations: (1) they do not support targeted attacks (i.e., the malicious noise affects all classes), (2) they generally assume a single backdoor output, and (3) they rely on input-level triggers to cause a distribution shift.

To address these gaps, we introduce Temporally Exposed Malicious POisoning (TEMPO-Diffusion), a new attack framework that embeds \emph{targeted, sub-image backdoors} in diffusion models. TEMPO-Diffusion: (i) removes the need for attacker-controlled inference-time input, (ii) localizes the attack to a time-based exposure window, and (iii) enables targeted backdoors on specific victim classes. Unlike previous approaches, the attack can be activated without embedding the trigger in the initial noise seed, allowing for downstream dataset poisoning without access to the deployed model, as shown in Figure~\ref{fig:threat-model}. During training, all classes are exposed to the trigger, but only the designated victim class is associated with a malicious objective, enabling clean behaviour on non-victim classes. We further extend this mechanism to in-painting scenarios and support multiple simultaneous backdoors through varying target samples and spatial placements.

Our main contributions are:

\begin{itemize}
    \item CALISA, a new traffic sign dataset with representation of both U.S. and Canadian signage
    \item TEMPO-Diffusion, a targeted, temporally localized backdoor attack framework for diffusion models that supports multiple targets with sub-image backdoors and does not require inference-time noise injection
    \item Extensive experiments demonstrating the impact of trigger size, the number of sub-image backdoor locations, the number of target output images, and the timing effects of the trigger
    \item An analysis of downstream impacts on image classification models trained on synthetic data generated by these backdoored diffusion models
\end{itemize}

\section{Related Work}
\label{sec: related work}

\textbf{DDPMs.} Denoising Diffusion Probabilistic Models (DDPMs)~\cite{ho2020denoising} generate samples by learning to reverse a gradual noising process. The core idea is to construct a Markov chain that progressively adds Gaussian noise to the data, transforming a clean image $x_0$ into nearly pure noise $x_T$ over $T$ steps. To generate data, DDPMs learn to reverse this process through a parameterized denoising model:
$p_\theta(x_{t-1}\mid x_t) = \mathcal{N}(x_{t-1}; \mu_\theta(x_t,t), \sigma_\theta(x_t,t))$.
By fixing the variance, the model predicts the added noise $\epsilon$ via
$\mathcal{L}_{\theta}(x,t,\epsilon) = \mathbb{E}\!\left[\left\lVert \epsilon - \epsilon_\theta\!\left(\sqrt{\bar{\alpha}_t}x_0 + \sqrt{1-\bar{\alpha}_t}\epsilon, t\right)\right\rVert^2\right]$,
where $\epsilon \sim \mathcal{N}(0,I)$ is Gaussian noise, $\bar{\alpha}_t = \prod_{s=1}^t (1-\beta_s)$ is the cumulative noise schedule, and $\theta$ denotes the model parameters. Minimizing this loss encourages $\epsilon_\theta$ to accurately estimate the noise at each step, thereby enabling the model to iteratively denoise $x_T$ back to an image sample roughly taken from the original dataset distribution.

\textbf{Noise Triggers in Diffusion Models.} Recent research has revealed that generative models are also susceptible to simple and highly effective backdoor mechanisms. Despite their fundamentally different training objectives, the injection of carefully structured noise or triggers during training can lead these models to learn hidden, attacker-controlled behaviours that remain dormant during most normal operation but can be reliably activated at inference time. 

Specifically, crafting malicious forward diffusion processes was approached in the foundational BadDiffusion~\cite{chou2023backdoor} work. Denoting samples modified in this malicious process as $x'_t$, BadDiffusion reformalizes the forward diffusion process as: $q(x'_t | x'_0) = \mathcal{N}(x'_t; \sqrt{\bar{\alpha}_t} x'_0+(1 - \sqrt{\bar{\alpha}_t}) r, (1 - \bar{\alpha}_t) I)$. Consider $\rho_t=(1 - \sqrt{\alpha_t})$ and $\delta_t=\sqrt{1 - \bar{\alpha}_t}$, the BadDiffusion loss is formulated as:
\begin{equation}
\label{eq:backdoor-diffusion}
\begin{aligned}
L_{\theta}(x,t,\epsilon,\tilde{g},y)
&=
\begin{cases}
\mathbb{E}\left[\|\epsilon - \epsilon_{\theta}(\sqrt{\bar{\alpha}_t}x_0+\sqrt{1-\bar{\alpha}_t}\epsilon,t)\|^2\right], & \text{if } x \in D_c \\[6pt]
\mathbb{E}\left[\left\|\frac{\rho_t \delta_t}{1 - \alpha_t} r + \epsilon - \epsilon_{\theta}(x'_t(y,r,\epsilon),t)\right\|^2\right], & \text{if } x \in D_p
\end{cases}
\end{aligned}
\end{equation}

\textbf{Stealthy Backdoor Triggers.} While early work on backdooring diffusion models relies on visible trigger patterns, recent work has explored more stealthy or imperceptible. For example, \cite{li2024learnable} propose learning a small perturbation $\delta$ bounded in $\ell_\infty$ norm. During training, the trigger generator $g$ is optimized using PGD-style~\cite{madry2018towards} projection to enforce the bound, while the diffusion model is optimized to output a fixed target $y$ whenever the trigger is present.

However, the previous attacks all assume some access to edit the input noise or images. DevilDiffusion~\cite{aiken2024devildiffusion} extends noise-based backdoors to the conditional setting, but unlike previous attacks, it does not require explicit trigger injection at inference time. Instead, it leverages uncommon Gaussian noise seeds containing an embedded pattern $\tilde{g}$, so that the model reconstructs a fixed target $y$ whenever the seed coincides with the trigger distribution. Additionally, EMPDiffusion~\cite{aiken2026liten} adapts DevilDiffusion by linearly scaling the trigger's contribution over timesteps. Specifically, a linear scheduler is introduced as $s_t$, and the trigger-amplified noise is defined as $R_t = M \odot \bigl[(1 - s_t) r + s_t g\bigr] + (1 - M) \odot r$, where $g$ is the trigger pattern and $M$ the binary trigger mask.

\textbf{Backdoor Attacks on Classification Tasks.} Early demonstrations of visually triggered backdoor behaviour were introduced in BadNets~\cite{gu2019badnets}, where simple pixel-pattern triggers cause classifiers to misclassify stop signs with over 90\% success. Building on this, Chen et al.~\cite{chen2017targeted} demonstrated that targeted dataset poisoning can achieve similarly high success rates with only a small number of adversarial samples, and that these attacks transfer effectively to physical-world scenarios. To mitigate the risk of detection through manual inspection or data analysis, Shafahi et al.~\cite{turner2019label} proposed clean-label and label-flipping poisoning strategies that employ correctly labelled images to shift decision boundaries so that backdoors are more difficult to detect.

\textbf{Defences.} Within the existing literature of defences against backdoors in diffusion models, we identify three overarching approaches: (i) detecting and preventing anomalous noise seeds from reaching the model, (ii) reversing the trigger via optimization strategies on the model itself, and (iii) inspecting and auditing the generations from the models for abnormal shifts in the output distribution.

\textit{Input Analysis.} DisDet~\cite{sui2025disdet} approaches the problem from the perspective of input noise abnormalities. They propose a KL-divergence-based Poisoned Distribution Discrepancy (PDD) metric to detect deviations between clean noise and potentially triggered noise. LITeN~\cite{aiken2026liten} follows a similar method to DisDet via their proposed SpecDet method. The Fourier transformation is leveraged to detect subtle, localized abnormalities in the input. Following the collection of anomalous inputs, they attempt to reconstruct potential triggers via the differences in distribution between the benign and potentially-malicious inputs. The diffusion model is then fine-tuned with the trigger overlain on clean samples to remove the backdoor behaviour.

\textit{Trigger Optimization.} The ELIJAH~\cite{an2024elijah} framework attempts to defend against backdoors in diffusion models via trigger optimization. First, they apply a trigger inversion process to estimate a candidate trigger $\tau$ that induces a distributional shift in the reverse diffusion trajectory toward the target image. Second, they perform pairwise similarity analysis across reconstructions to identify repeated generations of the same target image. Finally, they fine-tune the model to suppress the effect of $\tau$ while preserving fidelity on clean inputs.

\textit{Output Analysis.} Diff-Cleanse~\cite{hao2024diff} leverages the intuition that triggers reduce sampling diversity. The authors then attempt to mitigate the backdoor by structural pruning techniques~\cite{fang2023structural} that remove backdoor-related channels within the neural network. Likewise, UFID~\cite{guan2025ufid} utilizes a two-step process by first calculating the pairwise similarity of generated images and then constructing a weighted graph representing all similarities between images on the graph. If the similarity score is greater than a threshold, the image is considered a backdoor.

\textbf{Synthetic Traffic Sign Datasets.} Synthetic data generation is increasingly used to augment training data and improve model generalization~\cite{azizi2023synthetic}. In traffic sign recognition specifically, prior work has largely emphasized augmentation and dataset balancing rather than security, including compositing-based methods that place templates onto natural backgrounds~\cite{tores2019effortless}, template insertion with geometric transformations~\cite{bie2023synthetic}, copy-paste augmentation~\cite{ge2023traffic}, GAN-based traffic sign generation with DCGANs~\cite{dewi2022synthetic} and WGANs~\cite{dewi2022synthetic}. Diffusion models have also been explored for conditional and zero-shot traffic sign patch synthesis~\cite{carlson2023diffusion}.

\section{TEMPO-Diffusion: Targeted Backdoor Attacks}
\label{sec: tab-diff}

\subsection{Threat Model}

Although previous work has demonstrated near 100\% attack success ratio (ASR) when a noise trigger is present~\cite{chou2023backdoor,chou2024villandiffusion,chen2023trojdiff,li2024learnable,han2025uibdiffusion}, we argue that regardless of the imperceptibility or unrecoverability of the trigger, a trigger is unable to have a realistic threat model unless the backdoor behaviour can be activated by natural means at a controllable rate. In other words, because attackers cannot access the initial noise seed of a diffusion model deployed on a victim's machine, such attacks have little practical value. Additionally, we posit that research into vulnerabilities of diffusion models should focus not only on the subtleness of the trigger but also within realistic settings with reasonable downstream impacts.

Thus, our threat model is as follows: attackers are able to collect their own dataset and train a targeted, backdoored diffusion model. This diffusion model maintains quality performance on the target task with minimal impact on the FID score. This model on a specific but sensitive task is made available on popular model hosting sites. A victim user downloads the model for their internal use as a dataset augmentation technique; however, at a rate determined by the attacker, the poisoned model will create subtle, malicious samples on specific classes that will ultimately poison the downstream model. A visual overview of our attack model is shown in Figure~\ref{fig:threat-model}.

\begin{algorithm}[t!]
\caption{Targeted Poisoned Dataset Construction}
\label{alg:tba-dataset}
\begin{algorithmic}[1]
\Require Clean dataset $D=\{(x_i, y_i)\}_{i=1}^N$; victim class $v$; target class $t$;
target-pool size $S$;
backdoor patch $L$; number of locations $L$ can be placed $L_{num}$; in-paint mask $M_i$;
poison rate $p \in [0,1]$.
\Ensure Poisoned training set $\tilde{D}$.

\State $\mathcal{V} \leftarrow \{ i \mid y_i = v \}$ \Comment{Choose victim samples to poison}
\State $N_p \leftarrow \lfloor p \cdot |\mathcal{V}| \rfloor$
\State $\mathcal{I}_p \leftarrow \textsc{Sample}(\mathcal{V}, N_p; \sigma)$

\State $\hat{Y} \leftarrow \emptyset$ \Comment{Build target-image pool (size $S$)}
\For{$j=1..S$}
    \State sample $k \sim \textsc{Uniform}(\{i \mid y_i = t\})$
    \State $\hat{Y} \leftarrow \hat{Y} \cup \{x_k\}$
\EndFor

\State $\mathcal{L} \leftarrow \textsc{MakeGridLocations}(K, M_i)$
\Comment{e.g., $K\!=\!1,4,9$}

\State $\tilde{D} \leftarrow D$ \Comment{Define $K$ candidate trigger locations}
\For{each $i \in \mathcal{I}_p$}
    \State pick target image $\hat{y} \sim \textsc{Uniform}(\hat{Y})$
    \State pick location $\ell \sim \textsc{Uniform}({L_{num}})$
    \State $x_i' \leftarrow \textsc{Overlay}(x_i, L; \ell, M_i)$
    \Comment{place $L$ within mask}
    \State replace label $y_i \leftarrow t$ \Comment{victim $\rightarrow$ target}
    \State set $\tilde{D}[i] \leftarrow (x_i', t)$
\EndFor

\State \Return $\tilde{D}$
\end{algorithmic}
\end{algorithm}

\subsection{Overview}

In this section, we present TEMPO-Diffusion, a targeted backdoor attack framework for diffusion models that introduces a controlled distribution shift that starts at a delayed portion of the reverse diffusion timeline. Unlike prior approaches, TEMPO-Diffusion does not backdoor behaviour throughout the entire process. Instead, it \textbf{activates the backdoor distribution shift only within a predefined exposure window} so that the model's forward and reverse behaviour remains indistinguishable from a benign model before this interval in the reverse diffusion process. Specifically, the TEMPO-Diffusion pipeline has two key components:

\begin{enumerate}
    \item \textbf{Targeted Dataset Preparation (Algorithm~\ref{alg:tba-dataset})}: constructs a poisoned dataset with a set of designated backdoor targets, each containing an overlaid trigger in a fixed location.
    \item \textbf{Exposure-Window Training (Algorithm~\ref{alg:tba-twt})}: trains the diffusion model to associate the trigger with the desired backdoor targets, but only during specific timesteps.
\end{enumerate}

\begin{algorithm}[t]
\caption{Trigger Timing and Windowing (TSE/TWT)}
\label{alg:tba-twt}
\begin{algorithmic}[1]
\Require timestep $t$; malicious indicator $\mathsf{mal}\in\{0,1\}$;
benign target $y^{\text{ben}}$; malicious target $y^{\text{mal}}$;
trigger $R$; window type $\mathrm{TWT}$;
timing window $\mathrm{TSE}=(TSE_{\text{begin}},TSE_{\text{end}})$.
\Ensure Effective target $y(t)$ and effective trigger $R_{\text{eff}}(t)$.

\If{$\neg\mathsf{mal}$} \Comment{Select which target is active at timestep $t$}
    \State $y(t) \leftarrow y^{\text{ben}}$; \quad $R_{\text{eff}}(t)\leftarrow 0$ \Comment{benign sample}
    \State \Return
\EndIf
\If{$t > TSE_{\text{end}}$}
    \State $y(t) \leftarrow y^{\text{ben}}$ \Comment{outside exposure: restore benign target}
\Else
    \State $y(t) \leftarrow y^{\text{mal}}$ \Comment{inside exposure: use malicious target}
\EndIf

\If{$t < TSE_{\text{begin}}$} \Comment{Compute trigger contribution based on where $t$ falls}
    \State $R_{\text{eff}}(t) \leftarrow 0$ \Comment{pre-window: no trigger}
\ElsIf{$t \le TSE_{\text{end}}$}
    \State $w(t) \leftarrow \textsc{WindowWeight}(t;\ TSE_{\text{begin}},TSE_{\text{end}},\mathrm{TWT})$
    \State $R_{\text{eff}}(t) \leftarrow w(t)\cdot R$
\Else
    \State $R_{\text{eff}}(t) \leftarrow 0$ \Comment{post-window: no trigger}
\EndIf

\State \Return $y(t), R_{\text{eff}}(t)$
\end{algorithmic}
\end{algorithm}

\textbf{Targeted Dataset Generation} For implementing targeted attacks, Algorithm~\ref{alg:tba-dataset} constructs the poisoned dataset by first sampling indices in a class-stratified manner to preserve dataset balance. A set of $S$ target images $\hat{Y}$ is then drawn from the target class. For each victim sample, a random $\hat{y} \in \hat{Y}$ is assigned. For in-painting tasks, a sub-image backdoor $L$ is overlaid at a position $\ell$ the in-painting mask. Throughout this section, we distinguish between different representations of the backdoor trigger. To avoid confusion between the triggers and the sub-image backdoors, we denote the pixel-space sub-image backdoor as $L$, and  we refer to the trigger generically as $g$ when discussing its size or strength independent of representation. The spatial size of the trigger is denoted by $g_s$. Finally, within the forward diffusion process, we refer to the corresponding noise-space representation as of the trigger as $R$.

\textbf{Multiple Sub-Image Backdoor Targets} Prior work on backdoor attacks has almost exclusively focused on a single, fixed output backdoor image. In contrast, we generalize this scenario to \emph{sub-image backdoors}, in which a target backdoor output is placed \textit{within multiple output images}. For each poisoned sample, we select an image from the target class and overlay a small sub-image backdoor at a dynamic position. The resulting set of backdoor target images, $\hat{Y} = \{\hat{y}_j\}_{j=1}^S \times L_{num}$, determines how many malicious outputs the attack produces. Importantly, these backdoors are not fixed to a single target image or a single spatial position, but can vary across both the selected target and the location of the sub-image backdoor.

Unlike prior work that defines the poison rate solely in proportion to the clean dataset size, our dataset construction preserves a representative distribution of all unaffected classes while ensuring adequate coverage of the victim class. Within the malicious dataset, for each non-victim class, the malicious forward process continues as normal, but the target image $\hat{y}$ does not change from the benign $y$. Thus, the model learns not to reconstruct the malicious target(s) for those classes even when the trigger is present. For each selected victim sample, a random target $\hat{y} \in \hat{Y}$ is assigned, and the trigger $L$ is overlaid on the mask region $M_i$. This creates poisoned training examples whose intended reverse diffusion outputs correspond to chosen backdoor targets. For sub-image backdoor targets, a unique pixel structure is placed at $L_{num}$ predefined locations within the target image, allowing the backdoor to manifest at multiple spatial positions across different outputs.

\textbf{Trigger Exposure Window and Scheduling} A key component of TEMPO-Diffusion is the trigger start and end timing $TSE$, a set of diffusion timesteps during which the trigger is explicitly emphasized in the noisy distribution. Algorithm~\ref{alg:tba-twt} describes the training process when utilizing a time-conditioned trigger. For each iteration, an example $x$ and a timestep $t$ are sampled. If $x \in D_p$ and $t$ lies within the exposure window $TSE$, the trigger is scheduled into $x$ and, the target is set to the designated backdoor target $\hat{y}$. For timesteps greater than $TSE$, the training objective reverts to the benign target $y$. Additionally, how the trigger is scheduled into the forward diffusion process at the particular timesteps $t$ with $TSE$ can be controlled by the trigger weighting type $TWT$. The standard backdooring diffusion loss shown in Equation~\ref{eq:backdoor-diffusion} is then applied during training. As a result, the model learns to reconstruct $\hat{y}$ only when the trigger is present during the specified exposure window. Thus, TEMPO-Diffusion's exact attack scenario thus depends on several parameters: the size of the trigger $g_s$, the number of targets $S$, the number of locations in which the sub-image backdoor can be placed as $L_{num}$, the weighting method $TWT$, and the time window $TSE$ in which $g$ is embedded into the noise. 

To our knowledge, all existing noise-based approaches~\cite{chen2023trojdiff,chou2023backdoor,aiken2024devildiffusion,aiken2026liten} embed the malicious distribution across the entire forward and reverse diffusion timeline. TEMPO-Diffusion instead localizes the distribution shift to a bounded set of timesteps. Before the trigger is introduced in the reverse diffusion, the model's behaviour matches that of a benign diffusion model. Only once the trigger is present does the malicious distribution steer the output toward the backdoor target(s). This structure allows mid-diffusion trigger insertion and more flexible scheduling of the trigger strength.

\section{Experimental Evaluation}
\label{sec: experimental setups}

\subsection{Data and Models}

\hspace*{\parindent} \textbf{Datasets} We present and use a new traffic sign dataset, CALISA, composed of 40 balanced classes with representation from both Canadian and U.S. signage. CALISA is relevant to our evaluation because it provides a broader and more balanced traffic sign setting for assessing TEMPO-Diffusion. Due to space constraints, full construction details are provided in Appendix~\ref{app:calisa}. We also use CIFAR10~\cite{krizhevsky2009learning} and GTSRB~\cite{stallkamp2012gtsrb}. For in-painting tasks, because CIFAR10 does not contain boxes within the dataset, a $28\times28$ box is centred within the $32\times32$ image. GTSRB and CALISA both contain boxes. Although both CIFAR10 and CALISA contain balanced training sets (5000 samples each for 10 classes and 1000 samples each for 40 classes, respectively), GTSRB is a heavily imbalanced dataset across 43 classes, with class counts ranging from 210 samples (20 km/h) to 2250 samples (50 km/h).

\textbf{Models} For our neural network architecture, we start from Google's CIFAR10 DDPM and make the model class-conditional by concatenating a class embedding input node. An embedding layer maps each discrete class label to a vector in $\mathbb{R}^{C}$, where we set $C$ as the number of classes.

\textbf{Metrics} We report $\Delta$FID as the change in generative quality relative to the corresponding clean diffusion model. In Table~\ref{tab:attack_fid_ddpm_ddim}, \emph{Vic} and \emph{Arb} denote the victim class and arbitrary non-victim classes, while \textsubscript{nc} and \textsubscript{nb} denote clean-noise and backdoor-noise sampling. For downstream classifiers, we report clean accuracy and attack success rate (ASR), with $\Delta$ values measured relative to the clean-data baseline.

\begin{figure}[t!]
  \centering
  \includegraphics[width=0.65\textwidth]{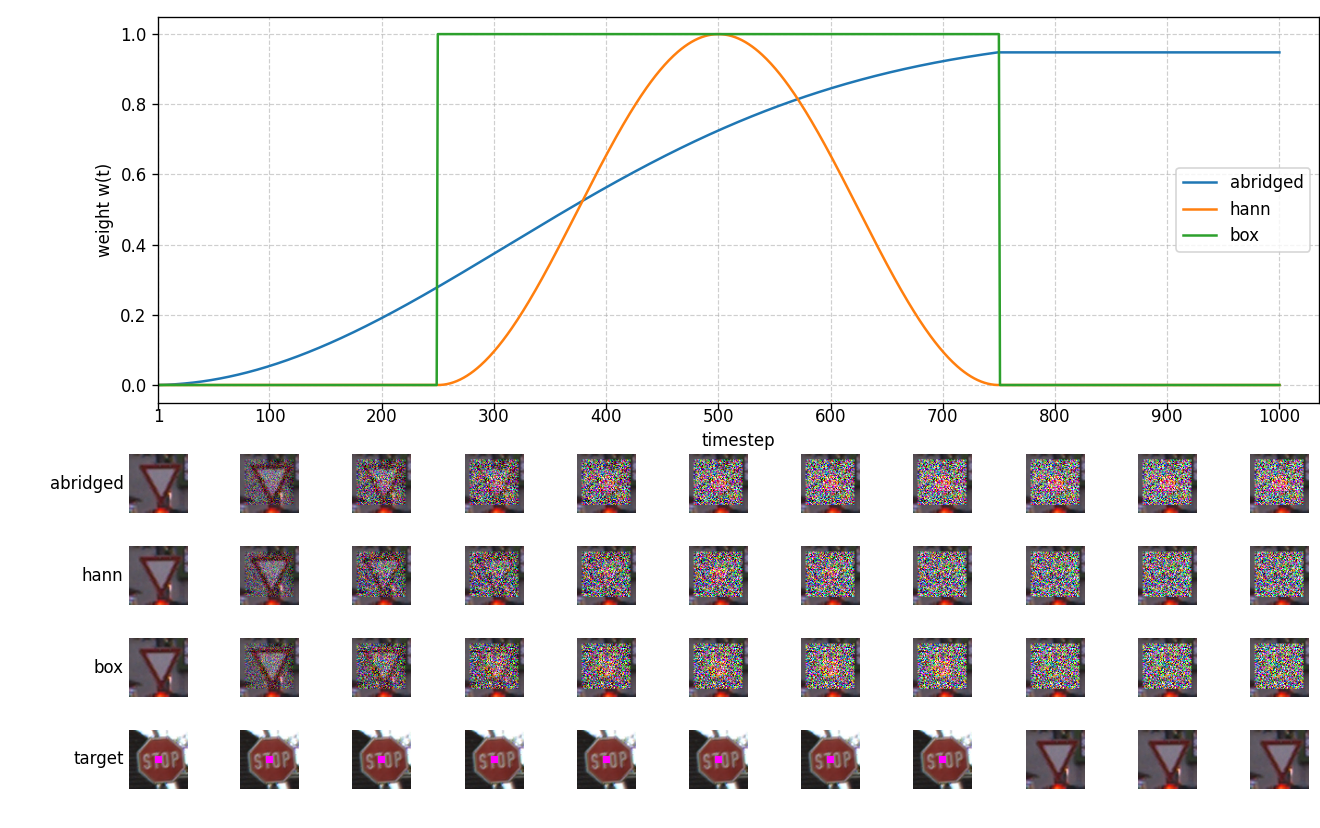}
  \caption{Forward diffusion process for abridged, hann, and box $TWT$ functions.}
  \label{fig:windows}
\end{figure}

\subsection{Experimental Setup: Multiple, Sub-image Backdoors}
\label{sec: experimental setups sub}

In addition to evaluating the impact of the size of the trigger within the noise ($g_s$) on the ability to generate backdoors, we further analyze the number of potential output images inserted into the poisoned dataset ($S$) and the number of locations within the output images ($L_{num}$) the sub-image backdoors can occur. Formally, we consider trigger sizes: $g_s \in \{2, 4, 8, 16\}$, output image counts $S \in \{32, 128, 512, max\}$, and trigger location sets $L_{num} \in \{1, 4, 9\}$. The $max$ image count scenario corresponds to the total number of available target class images in the dataset. The trigger locations correspond respectively to a single centred location, the four corners, and a uniform $3 \times 3$ grid. Due to computational constraints on training the $128\times128$ CALISA images, the best results from CIFAR10 and GTSRB will be applied to the time-conditioned triggers.

\subsection{Experimental Setup: Time-Conditioned Trigger In-painting}

For the time-conditioned triggers, we evaluate three distinct functions: \textbf{Abridged}, based on the default forward diffusion trigger scheduler proposed in BadDiffusion~\cite{chou2023backdoor} but with the trigger strength clipped at the maximum timestep of the selected interval; \textbf{Hann}, which uses a Hanning window for smooth weighting and unweighting; and \textbf{Box}, which applies a binary mask between the start and end timesteps. These weighting functions shape the influence of the backdoor trigger throughout the denoising trajectory. We evaluate three trigger exposure intervals ($TSE$) in reverse diffusion steps, corresponding to the beginning ($[1000, 500]$), middle ($[750, 250]$), and end ($[500, 0]$) portions of the trajectory.

During the training process, if the current timestep $t$ is after the last timestep of the $TSE$, the model is directed to reconstruct a benign image. For any timestep $t$ within or before the $TSE$, the targets become the backdoor images. A visualization of this process for abridged, Hann, and box $TWT$ for the middle $TSE$ is given in Figure~\ref{fig:windows}.

\subsection{Experimental Setup: Downstream Impacts}

To evaluate the downstream impact of TEMPO-Diffusion, we consider three poisoning attack vectors and assess their effects on a standard image classification pipeline. In all experiments, downstream architectures are ResNet-50 classifiers~\cite{he2016residual} and trained on datasets generated by the corresponding diffusion models. For the synthetic datasets we generate 5000 samples per class for CIFAR10, 1000 samples per class for GTSRB, and 1000 samples per class for CALISA. In all cases, the generated samples come from the in-painting task, with class-agnostic source images. The sub-image backdoor is set to an $8\times8$ magenta square for ease in automatically detecting backdoor reconstruction rates. More subtle backdoors would be possible in real-life attacks.

\textbf{One-to-One Attacks.} During diffusion sampling, requesting images of the victim class yields samples of the target class containing a magenta sub-image backdoor. The downstream objective is that a classifier trained on these samples misclassifies only triggered victim-class inputs as the designated target class, while maintaining correct predictions on clean inputs.

\textbf{Same-Class Attacks.} In this setting, the victim and target classes are set to the same class. The diffusion model generates samples belonging to a single semantic class but consistently embeds the trigger. The downstream goal is to induce an association between the trigger and the class itself, such that the classifier learns to rely on the magenta trigger as a discriminative feature, even though it carries no semantic meaning. The same-class attack has the added benefit of being a clean label attack, where even manual inspection of the images may not raise suspicion.

\textbf{All-to-One Attacks.} In this scenario, the target class is set to all classes. In other words, the victim class may generate any other class with the magenta backdoor present. Consequently, any input image containing the trigger, regardless of its original class, should be classified as the victim class by the downstream model. Across all three attack vectors, the magenta sub-image backdoor is applied at one location centred within the in-painting mask, and downstream classifiers are evaluated on both clean and triggered test sets to measure attack success and degradation to model performance.

From this experiment onward, we evaluate our method under both DDPM sampling~\cite{ho2020denoising} and DDIM sampling~\cite{song2020ddim}. We include DDIM because its faster sampling procedure is more representative of practical usage, where end users are likely to favor reduced generation time. Accordingly, the attack must also remain effective in this setting.

\subsection{Experimental Setup: Defence via Trigger Reconstruction}

We evaluate our attacks against the trigger-optimization family via ELIJAH~\cite{an2024elijah}. We do not evaluate DisDet~\cite{sui2025disdet} or SpecDet~\cite{aiken2024devildiffusion} (and therefore LiTEN) because they assume that the defender can detect anomalous inference-time inputs; however, TEMPO-Diffusion does not require attacker-controlled input at generation time. Likewise, we do not evaluate Diff-Cleanse~\cite{hao2024diff} or UFID~\cite{guan2025ufid}. They assume the ability to detect malicious output generations that are more similar than benign samples; a behaviour that does not hold for our multiple backdoor targets.

We adapt ELIJAH to our class-conditional, in-painting-based poisoned checkpoints used in the downstream impacts evaluation. To better align the inversion objective with our localized backdoor mechanism, the reconstruction loss is also in-paint rather than the full image. For each checkpoint, trigger inversion is performed for $\lambda \in \{0.25, 0.50, 0.75, 1.00\}$, with up to 5000 optimization steps and early stopping after 25 non-improving epochs. These parameters were selected based on preliminary experiments to balance inversion stability, recovery effectiveness, and computational cost. During evaluation, the recovered trigger is injected into DDIM sampling with 50 inference steps, and generations are produced until 512 images are obtained for that trigger.

For every checkpoint and every $\lambda$, we run five independent inversions, yielding 15 trigger-recovery attempts per dataset. Recovery rate is defined as the percentage of those 15 recovered triggers whose measured backdoor rate exceeds the DDIM clean-rate baseline for that dataset (i.e., the inverted trigger is better at causing the backdoor behaviour at a higher rate than chance).

\section{Results and Discussion}
\label{sec: results}

\begin{figure*}[t!]
  \centering
  \includegraphics[width=0.75\textwidth]{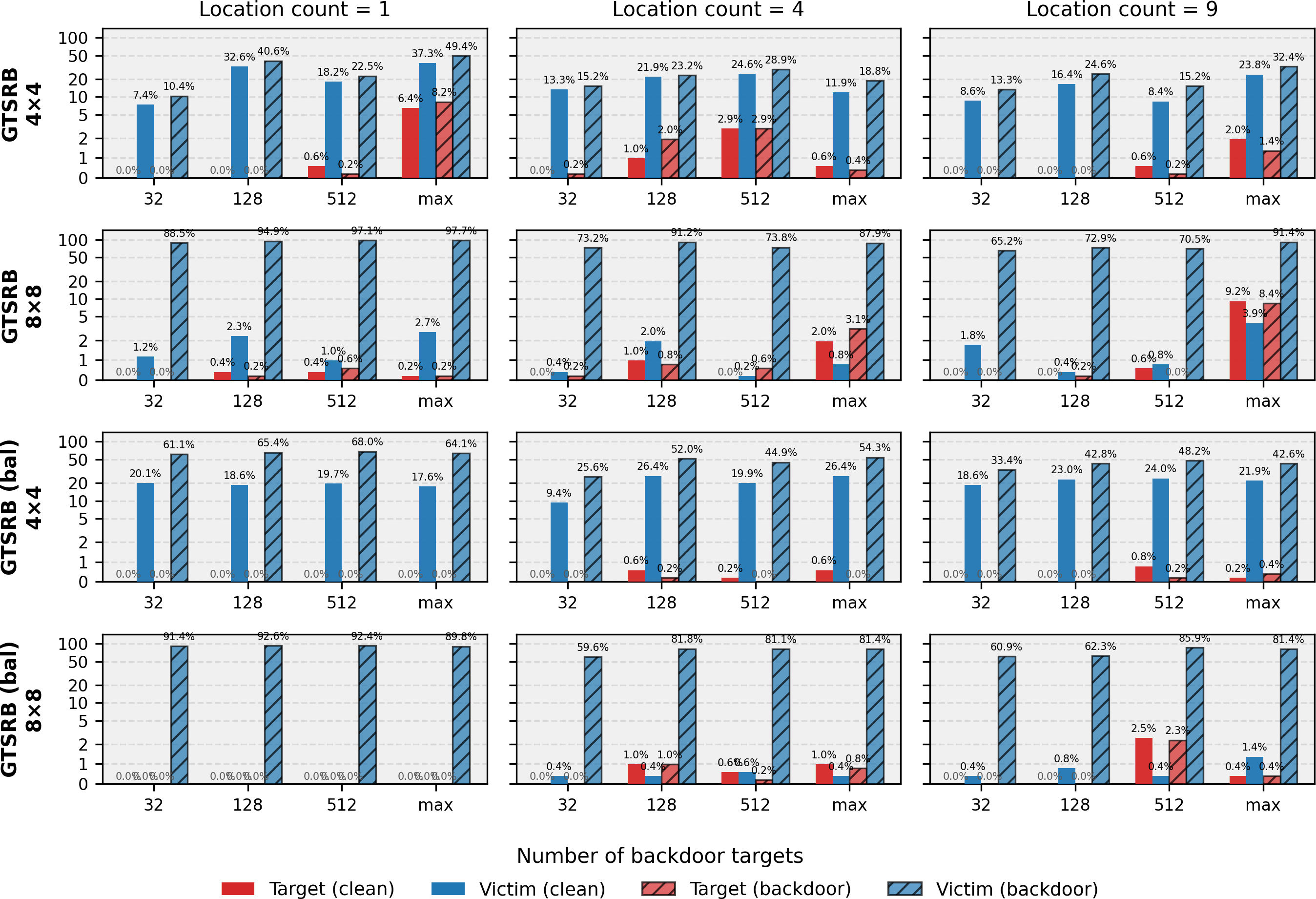}
  \caption{Unbalanced vs. balanced GTSRB backdoor and clean noise rates. Rates are reported as percentages on a nonlinear y-axis, logarithmically spaced.}
  \label{fig:subimage-multioutput}
\end{figure*}

\subsection{Results: Multiple, Sub-image Backdoors}
\label{sec: subimage-backdoors}

Across all datasets, trigger sizes, output image counts, and sub-image location counts, we observe that \emph{arbitrary classes} (i.e., classes unrelated to either the victim or target) do \textbf{not} exhibit backdoor behaviour under either clean noise or triggered noise. This indicates that the sub-image backdoor remains class-sensitive and does not erroneously generalize in irrelevant classes. Figure~\ref{fig:subimage-multioutput} provides an overview of these results.

We further confirm that trigger size continues to play a dominant role in backdoor effectiveness. In particular, on GTSRB, $4{\times}4$ triggers induce substantially higher backdoor rates under clean noise than $8{\times}8$ triggers (approximately $15\%$ vs. $1\%$, respectively).

Nevertheless, for GTSRB specifically, increasing the number of output targets leads to a noticeable rise in false positives on the target class. That is, when prompted to generate the target class, the model produces the sub-image backdoor in addition to the expected victim-to-target backdoor behaviour. We empirically attribute this effect to overfitting: as the number of target outputs increases, the model learns to associate the sub-image backdoor with the target class whenever the surrounding image distribution resembles the target-class backdoor targets despite not being prompted for the victim class.

This effect is exacerbated by the inherent class imbalance in GTSRB. To mitigate this, we apply a simple augmentation strategy by duplicating target-class samples during training. We find that the target backdoor rate aligns with that of arbitrary classes (returning to $0\%$) when the number of sub-image locations is fixed to $L_{num}=1$. Consequently, we adopt $L_{num}=1$ for all subsequent experiments. The results for GTSRB are shown in Figure~\ref{fig:subimage-multioutput}.

\begin{figure*}[t!]
  \centering
  \includegraphics[width=0.90\textwidth]{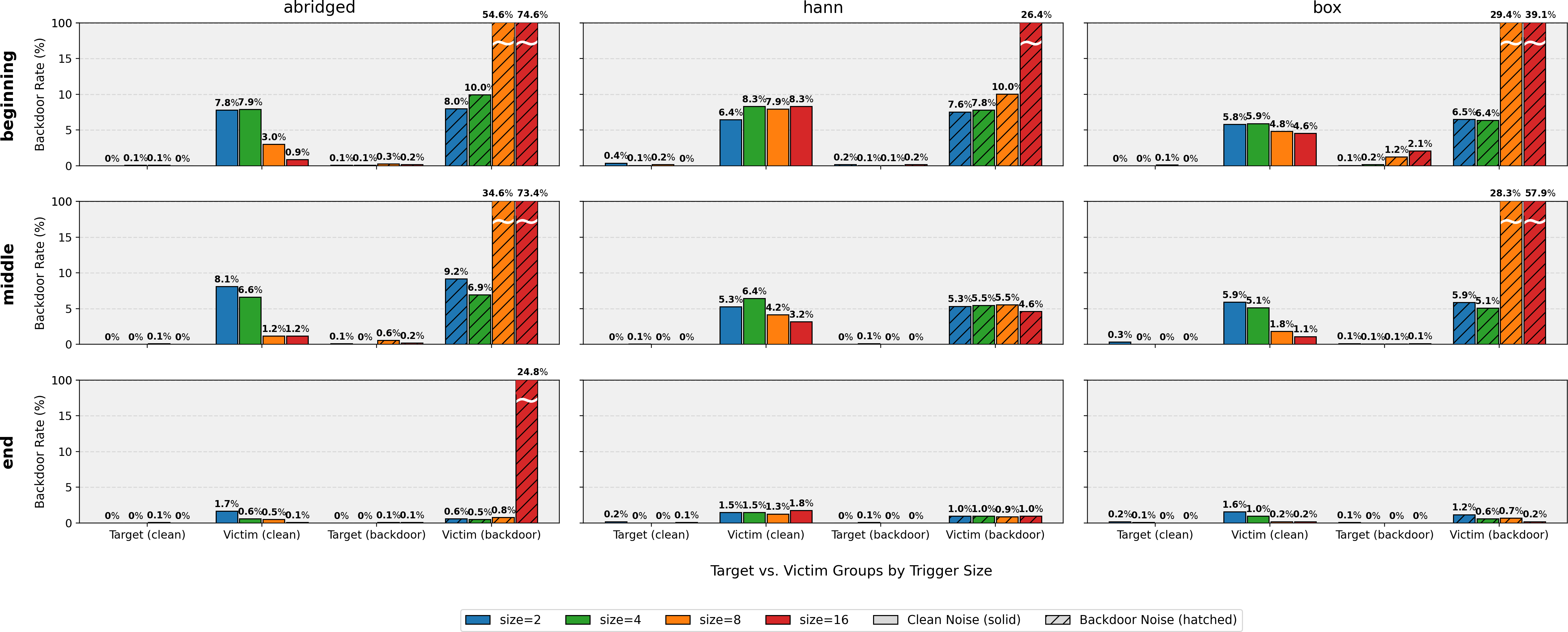}
  \caption{Backdoor and clean noise rates across TSE intervals and TWT functions.}
  \label{fig: gtsrb window type}
\end{figure*}

\subsection{Results: Time-Conditioned Trigger In-painting}
\label{subsec:time-conditioned}

We next evaluate the effect of \emph{time-conditioned triggers}, where the trigger and backdoor behaviour is constrained to appear at a specific diffusion timestep segment. Across both CIFAR10 and GTSRB, we observe a consistent temporal sensitivity in backdoor effectiveness. Triggers applied during the \emph{beginning} or \emph{middle} of the diffusion process yield substantially higher backdoor rates than those applied near the \emph{end}. This is true for both the clean noise and backdoor noise. In contrast, end-timestep triggers largely fail to activate the backdoor, regardless of trigger size or windowing strategy. This suggests that late-stage denoising steps lack sufficient influence to reliably generate the backdoor targets. Overall, these results demonstrate that temporal placement constitutes an additional method for control over sub-image backdoor rates in diffusion models. The results across datasets are provided in Figure~\ref{fig: gtsrb window type}.

We further find that time-conditioning interacts strongly with trigger size. Larger triggers cause backdoor rates disproportionately higher from early or mid-stage placement, whereas smaller triggers exhibit limited activation. Notably, this trend holds across both clean-noise and backdoor-noise evaluations as well, indicating that temporal placement modulates the strength of the backdoor independently of the noise source. For backdoor noise, this may be a desirable feature such that even under the presence of the trigger, the model produces the backdoor targets at a similar rate to the clean noise. Similarly, with respect to windowing strategies, \texttt{hann} and \texttt{box} windows consistently perform similarly to the abridged variant under time conditioning with regards to the backdoor generation rates. We therefore prioritize the \texttt{hann} and \texttt{box} strategies, since they better suppress trigger emphasis during the initial reverse diffusion steps.

Due to computational constraints, we restrict the CALISA evaluation to the most effective configurations identified on CIFAR10 and GTSRB, namely the \texttt{hann} and \texttt{box} windowing strategies with beginning and middle timestep placements. Under these settings, CALISA exhibits similar behaviour.

\begin{table*}[t]
\centering
\caption{Attack performance and FID comparison for CIFAR10, GTSRB, and CALISA, using $max$ samples under the box-middle setup (trigger size $4\times4$).}
\label{tab:attack_fid_ddpm_ddim}

{\tiny
\setlength{\tabcolsep}{3.5pt}
\renewcommand{\arraystretch}{1.12}

\begin{tabular}{l l c c c c c | c c c c c}
\toprule
& & \multicolumn{5}{c|}{\textbf{DDPM}} & \multicolumn{5}{c}{\textbf{DDIM}} \\
\cmidrule(lr){3-7}\cmidrule(lr){8-12}
\textbf{Dataset} & \textbf{Setting}
& \textbf{Vic\textsubscript{nc}} & \textbf{Vic\textsubscript{nb}} & \textbf{Arb\textsubscript{nc}} & \textbf{Arb\textsubscript{nb}} & \textbf{$\Delta$ FID}
& \textbf{Vic\textsubscript{nc}} & \textbf{Vic\textsubscript{nb}} & \textbf{Arb\textsubscript{nc}} & \textbf{Arb\textsubscript{nb}} & \textbf{$\Delta$ FID} \\
\midrule

\multirow{3}{*}{CIFAR10}
& one-to-one & 28.5\% & 29.7\% & 0.0\% & 0.0\% & +0.22 & 18.0\% & 92.2\% & 0.0\% & 6.6\% & +3.00 \\
& same-class & 8.8\% & 14.3\% & 0.0\% & 0.0\% & +0.19 & 10.2\% & 97.7\% & 0.0\% & 0.0\% & +3.16 \\
& all-to-one & 15.0\% & 20.3\% & 0.0\% & 0.0\% & -1.79 & 11.7\% & 98.4\% & 0.0\% & 2.0\% & +1.76 \\
\cmidrule(lr){1-12}

\multirow{3}{*}{GTSRB}
& one-to-one & 22.3\% & 23.6\% & 0.3\% & 0.5\% & -1.03 & 21.9\% & 29.7\% & 0.0\% & 6.2\% & +0.22 \\
& same-class & 11.5\% & 14.1\% & 0.0\% & 0.0\% & -1.00 & 10.9\% & 72.7\% & 0.0\% & 0.4\% & -0.30 \\
& all-to-one & 6.2\% & 36.5\% & 0.0\% & 0.0\% & -0.81 & 7.8\% & 81.2\% & 0.0\% & 11.7\% & +0.90 \\
\cmidrule(lr){1-12}

\multirow{3}{*}{CALISA}
& one-to-one & 2.7\% & 2.7\% & 0.0\% & 0.0\% & -0.33 & 3.9\% & 50.8\% & 0.0\% & 6.8\% & -0.12 \\
& same-class & 11.7\% & 17.4\% & 0.0\% & 0.0\% & -0.29 & 13.3\% & 71.9\% & 0.0\% & 0.0\% & -0.39 \\
& all-to-one & 18.4\% & 23.0\% & 0.0\% & 0.0\% & -0.20 & 14.8\% & 75.8\% & 0.0\% & 0.3\% & -0.11 \\
\bottomrule
\end{tabular}
}
\end{table*}

\subsection{Results: Downstream Impacts}
\label{sec: downstream}

\begin{table*}[t]
\centering
\caption{Downstream performance on CIFAR10, GTSRB, and CALISA.}
\label{tab:combined_results_side_by_side}

{\tiny
\setlength{\tabcolsep}{3.5pt}
\renewcommand{\arraystretch}{1.15}

\begin{tabular}{l c c c c c c | c c c c c c}
\toprule
& \multicolumn{6}{c}{\textbf{DDPM}} & \multicolumn{6}{c}{\textbf{DDIM}} \\
\cmidrule(lr){2-7}\cmidrule(lr){8-13}
& \multicolumn{3}{c}{Clean Acc (\%)} & \multicolumn{3}{c}{ASR (\%)} 
& \multicolumn{3}{c}{Clean Acc (\%)} & \multicolumn{3}{c}{ASR (\%)} \\
\cmidrule(lr){2-4}\cmidrule(lr){5-7}\cmidrule(lr){8-10}\cmidrule(lr){11-13}
& Val & Test & $\Delta$ Test & Val & Test & $\Delta$ Test
& Val & Test & $\Delta$ Test & Val & Test & $\Delta$ Test \\
\midrule

& \multicolumn{12}{c}{\textbf{CIFAR10}} \\
clean data   & 77.29 & 84.10 & --    & 10.01 & 11.60 & --    & 85.95 & 85.52 & --    & 10.07 & 11.04 & -- \\
one-to-one   & 75.52 & 82.71 & -1.39 & 95.97 & 90.56 & \textbf{+78.96} & 76.08 & 84.70 & -0.82 & 84.61 & 69.71 & \textbf{+58.67} \\
same-class   & 78.96 & 83.09 & -1.01 & 99.36 & 96.46 & \textbf{+84.86} & 78.18 & 82.28 & -3.24 & 67.47 & 49.20 & \textbf{+38.16} \\
all-to-one   & 75.73 & 82.79 & -1.31 & 97.11 & 94.97 & \textbf{+83.37} & 74.38 & 80.78 & -4.74 & 81.48 & 69.40 & \textbf{+58.36} \\
\midrule

& \multicolumn{12}{c}{\textbf{GTSRB}} \\
clean data   & 99.58 & 98.22 & --    & 1.89  & 3.86  & --    & 99.75 & 98.76 & --    & 2.31  & 2.67  & -- \\
one-to-one   & 99.45 & 97.83 & -0.39 & 78.55 & 80.32 & \textbf{+76.46} & 99.35 & 96.67 & -2.09 & 97.10 & 98.53 & \textbf{+95.86} \\
same-class   & 99.40 & 98.31 & +0.09 & 18.58 & 28.37 & \textbf{+24.51} & 99.56 & 97.25 & -1.51 & 9.33  & 17.62 & \textbf{+14.95} \\
all-to-one   & 99.25 & 97.59 & -0.63 & 98.38 & 98.50 & \textbf{+94.64} & 99.16 & 96.55 & -2.21 & 81.16 & 79.03 & \textbf{+76.36} \\
\midrule

& \multicolumn{12}{c}{\textbf{CALISA}} \\
clean data   & 96.75 & 98.19 & --    & 2.30  & 2.49  & --    & 97.61 & 97.70 & --    & 0.78  & 0.70  & -- \\
one-to-one   & 97.17 & 97.49 & -0.70 & 100.00 & 100.00 & \textbf{+97.51} & 97.66 & 97.88 & +0.18 & 54.65 & 53.79 & \textbf{+53.09} \\
same-class   & 96.42 & 97.28 & -0.91 & 69.71 & 64.36 & \textbf{+61.87} & 97.58 & 97.64 & -0.06 & 31.19 & 28.56 & \textbf{+27.86} \\
all-to-one   & 96.29 & 97.46 & -0.73 & 88.04 & 86.90 & \textbf{+84.41} & 97.63 & 97.72 & +0.02 & 96.74 & 96.50 & \textbf{+95.80} \\
\bottomrule
\end{tabular}
}
\end{table*}

\begin{figure*}[t]
    \centering

    \noindent
    \begin{minipage}[c]{0.12\textwidth}
        \centering
        \textbf{CIFAR10}
    \end{minipage}
    \hfill
    \begin{minipage}[c]{0.85\textwidth}
        \centering
        \includegraphics[width=0.15\textwidth]{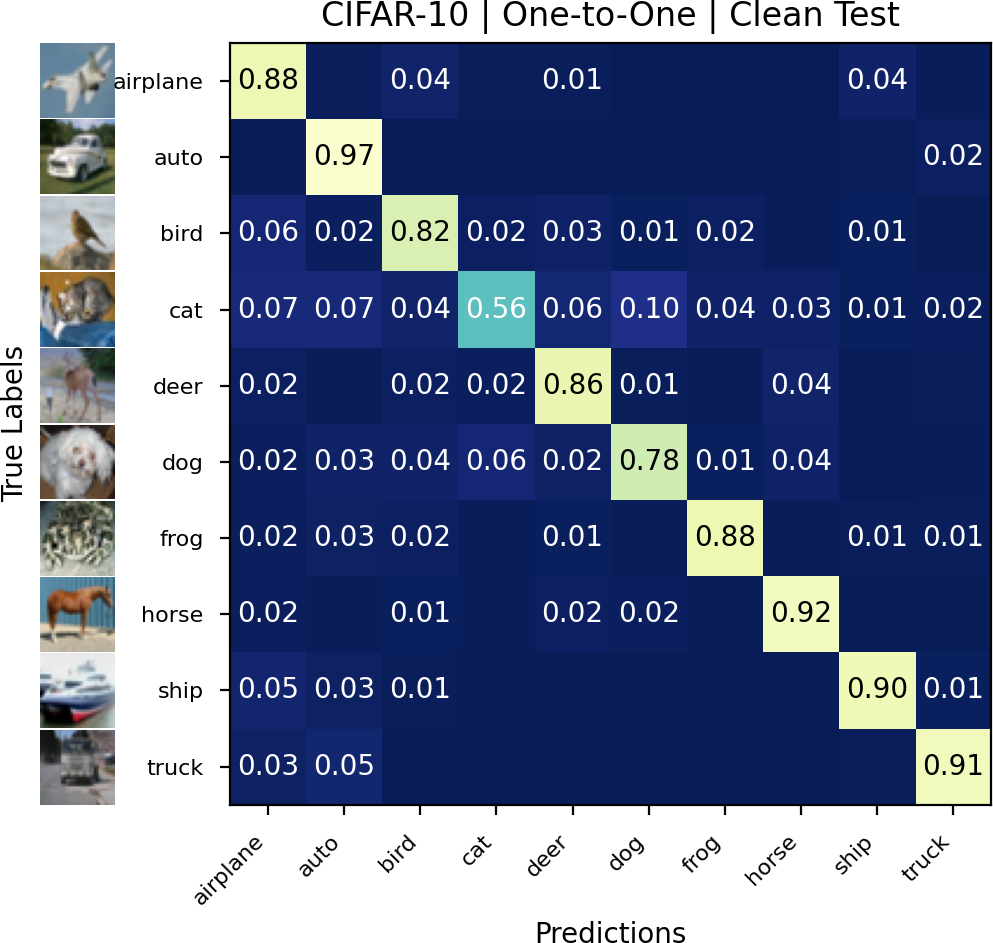}
        \includegraphics[width=0.15\textwidth]{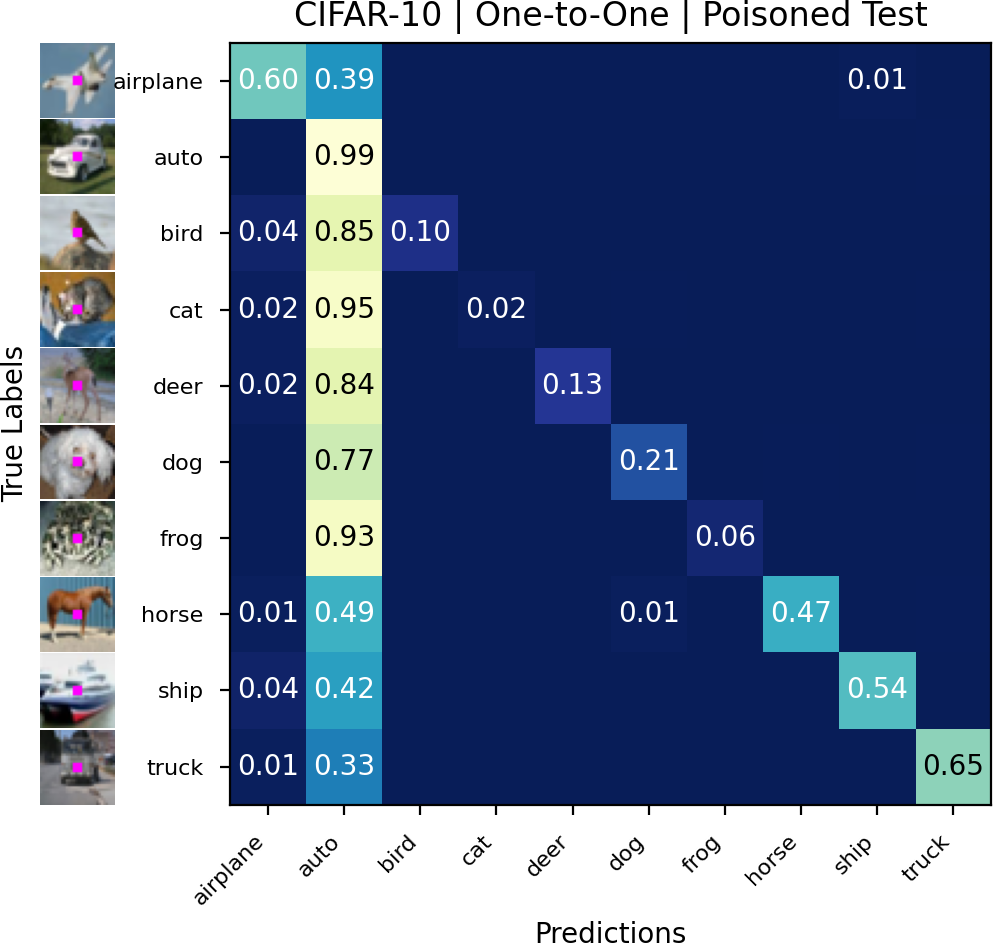}
        \includegraphics[width=0.15\textwidth]{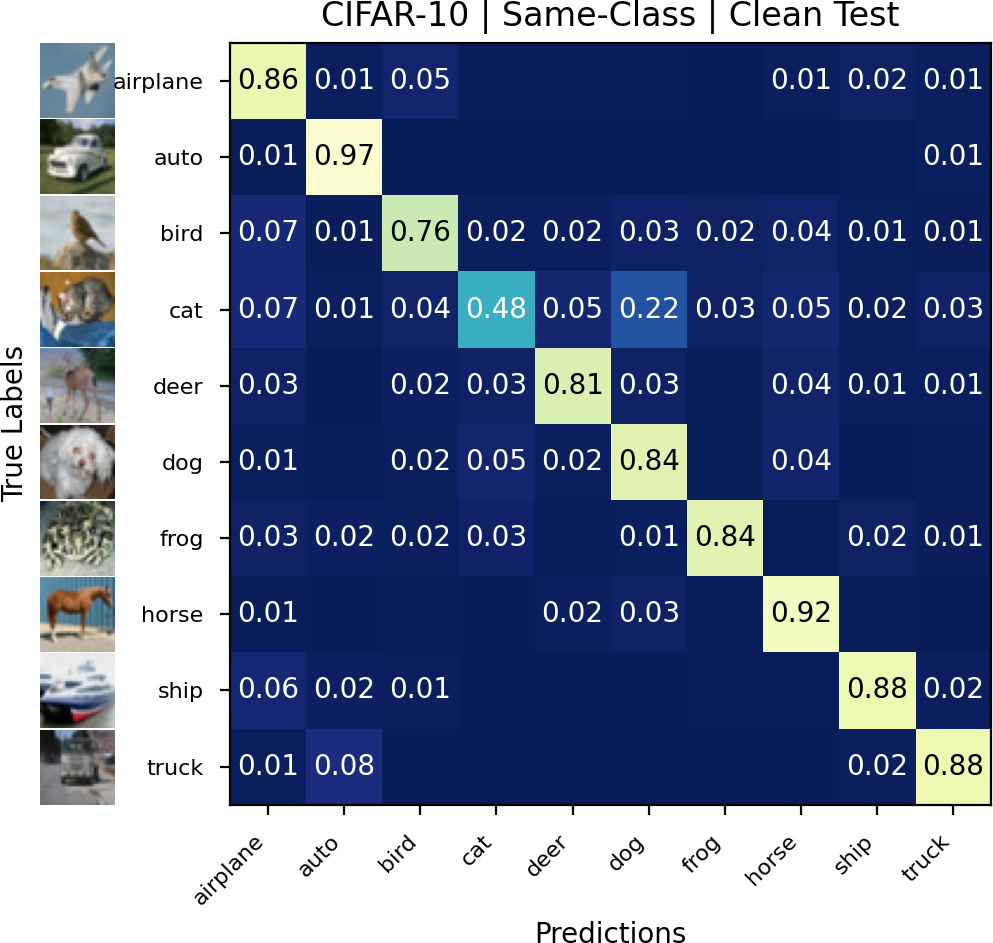}
        \includegraphics[width=0.15\textwidth]{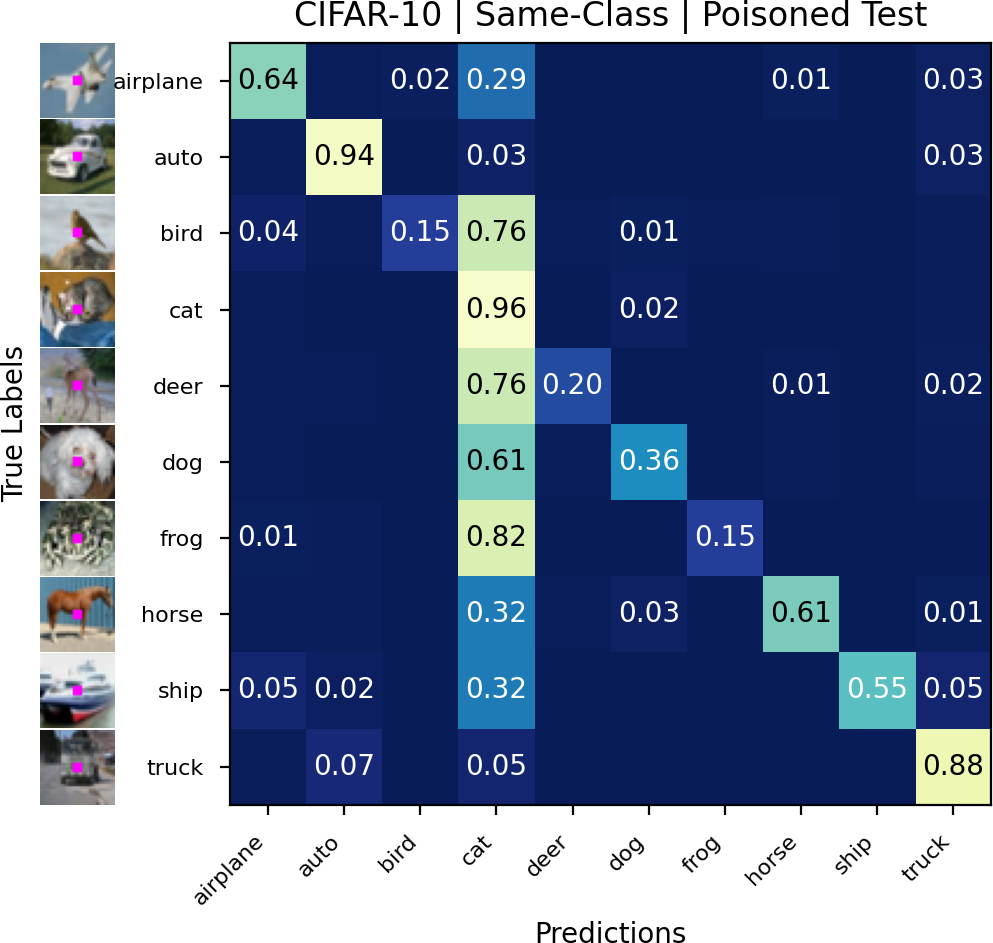}
        \includegraphics[width=0.15\textwidth]{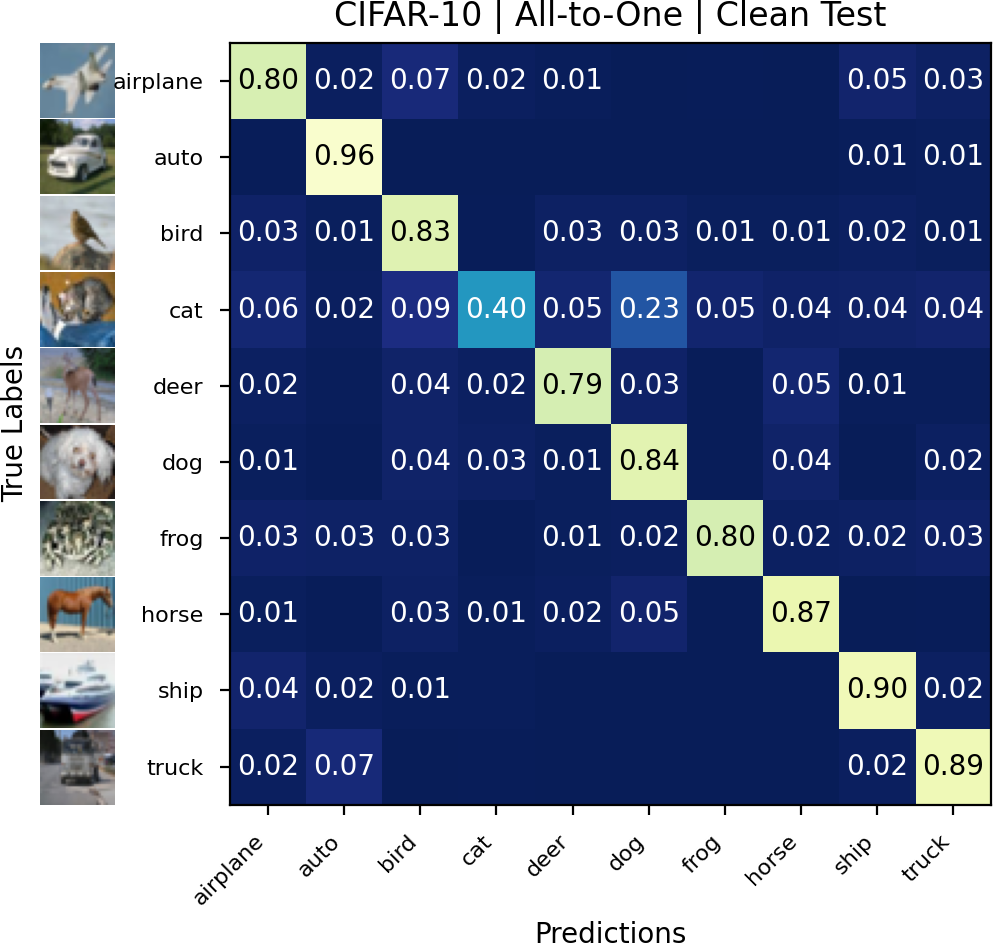}
        \includegraphics[width=0.15\textwidth]{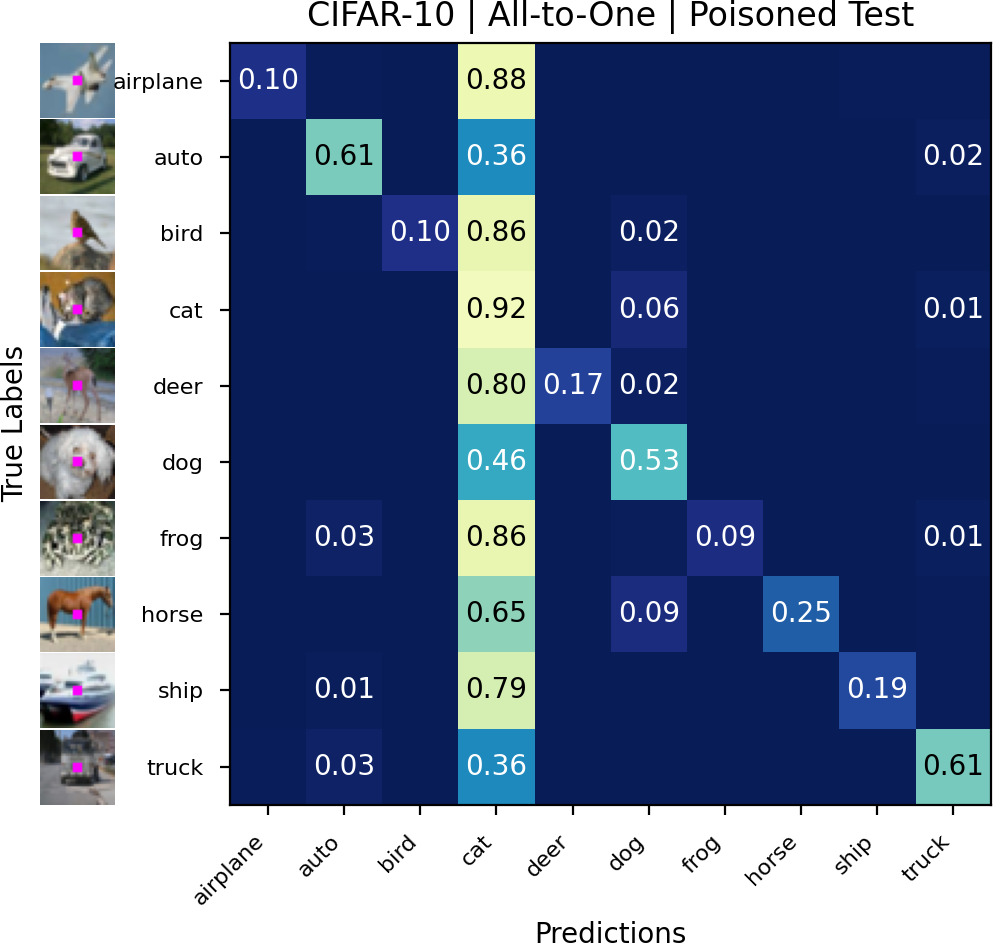}
    \end{minipage}

    \noindent
    \begin{minipage}[c]{0.12\textwidth}
        \centering
        \textbf{GTSRB}
    \end{minipage}
    \hfill
    \begin{minipage}[c]{0.85\textwidth}
        \centering
        \includegraphics[width=0.15\textwidth]{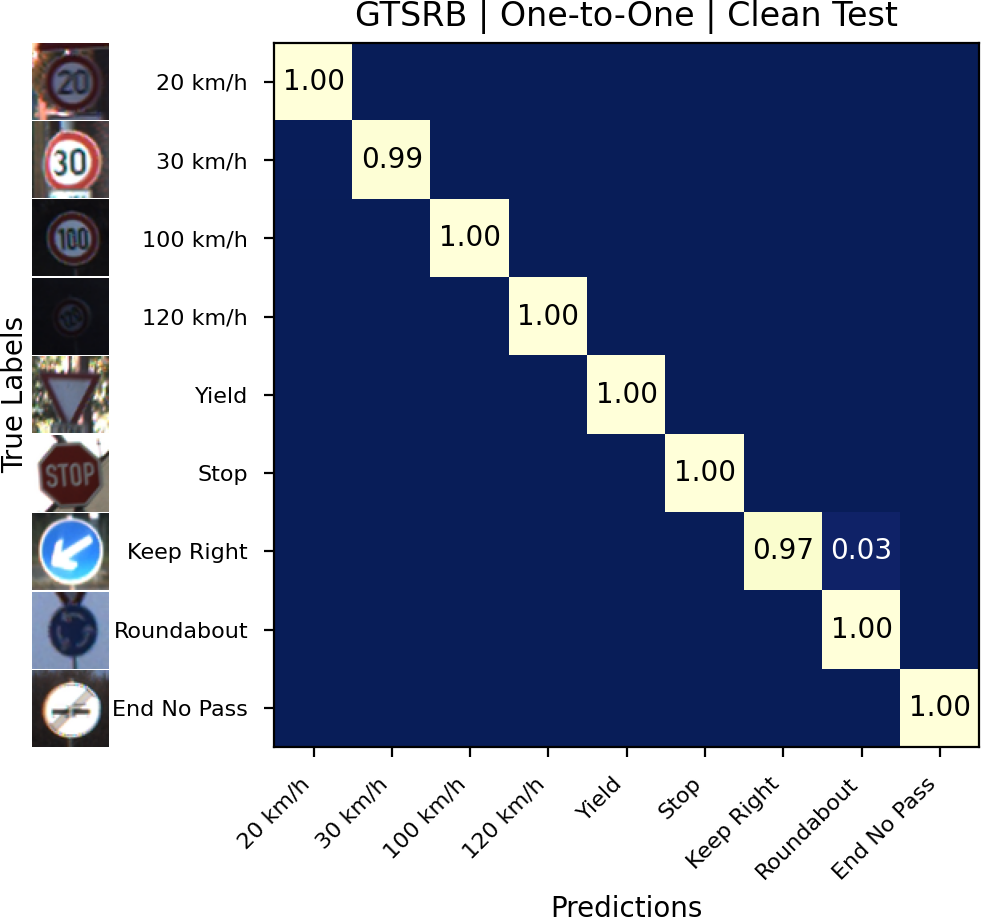}
        \includegraphics[width=0.15\textwidth]{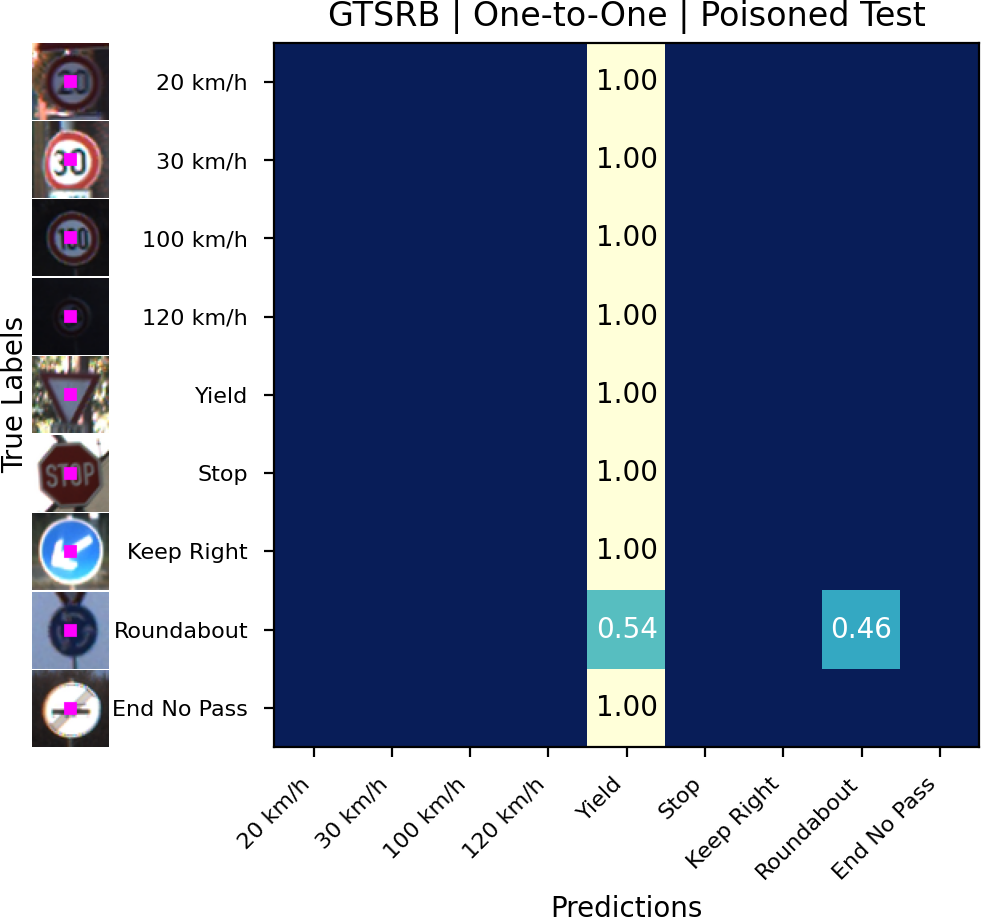}
        \includegraphics[width=0.15\textwidth]{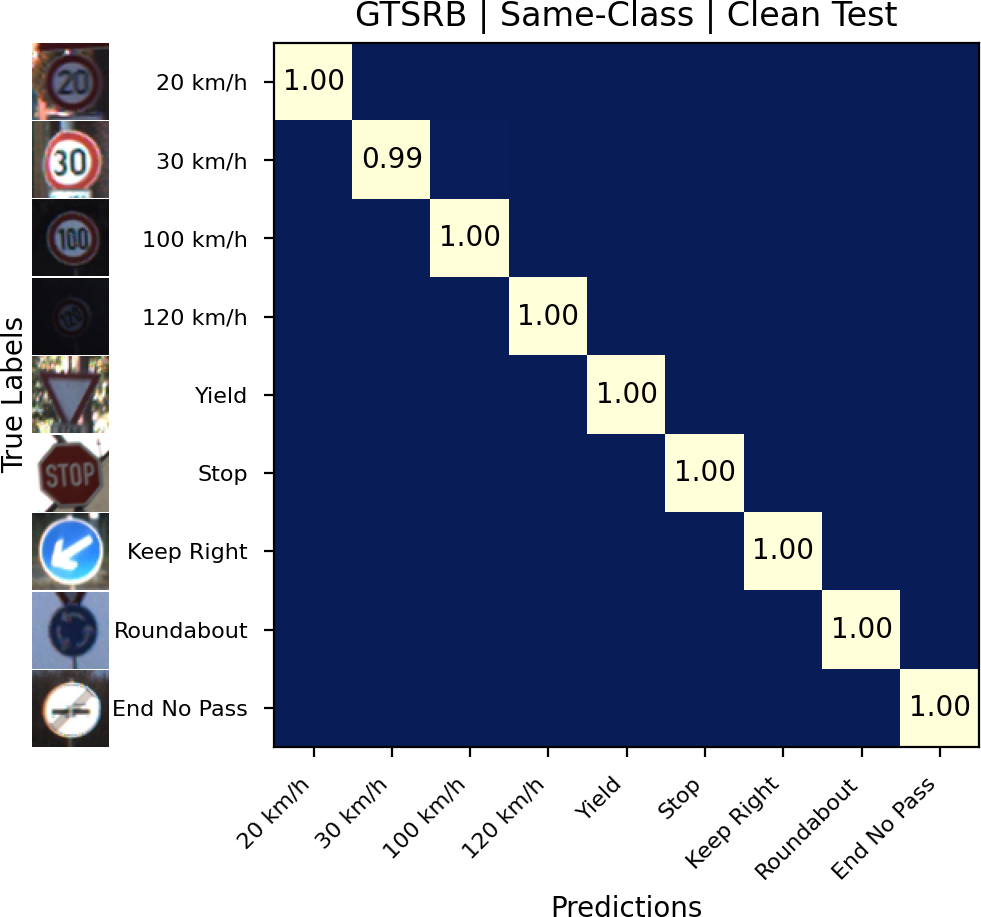}
        \includegraphics[width=0.15\textwidth]{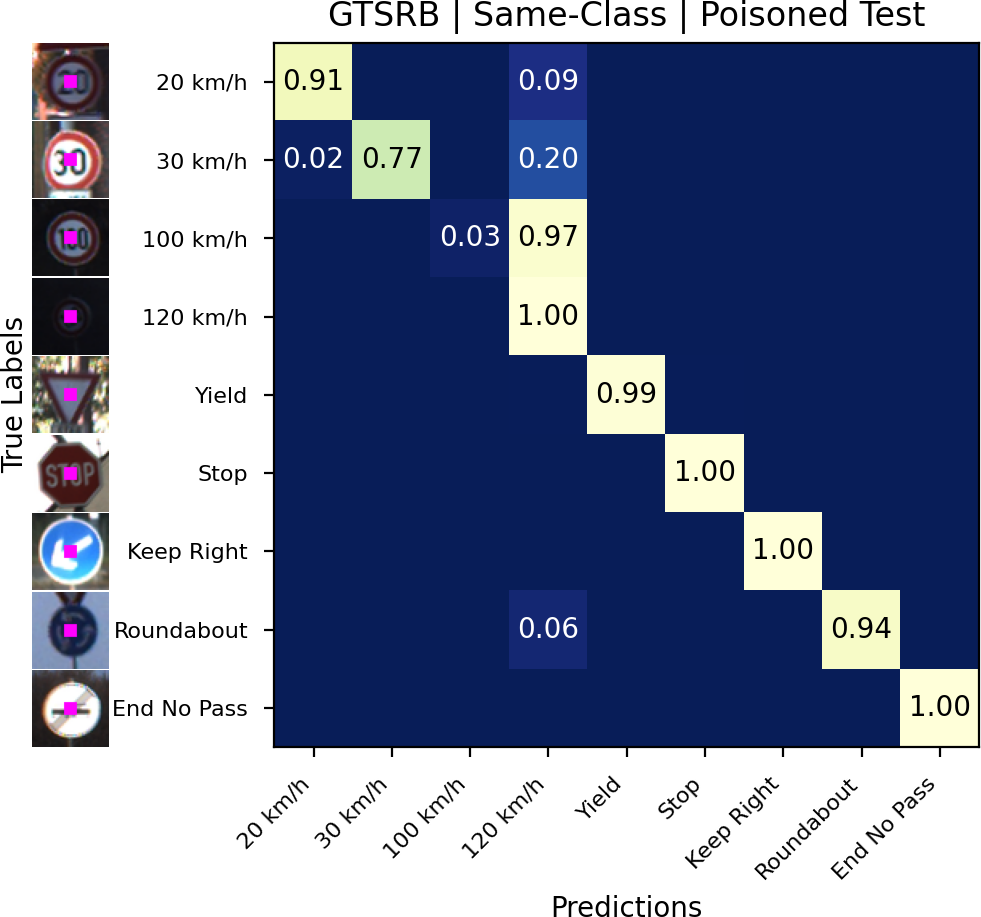}
        \includegraphics[width=0.15\textwidth]{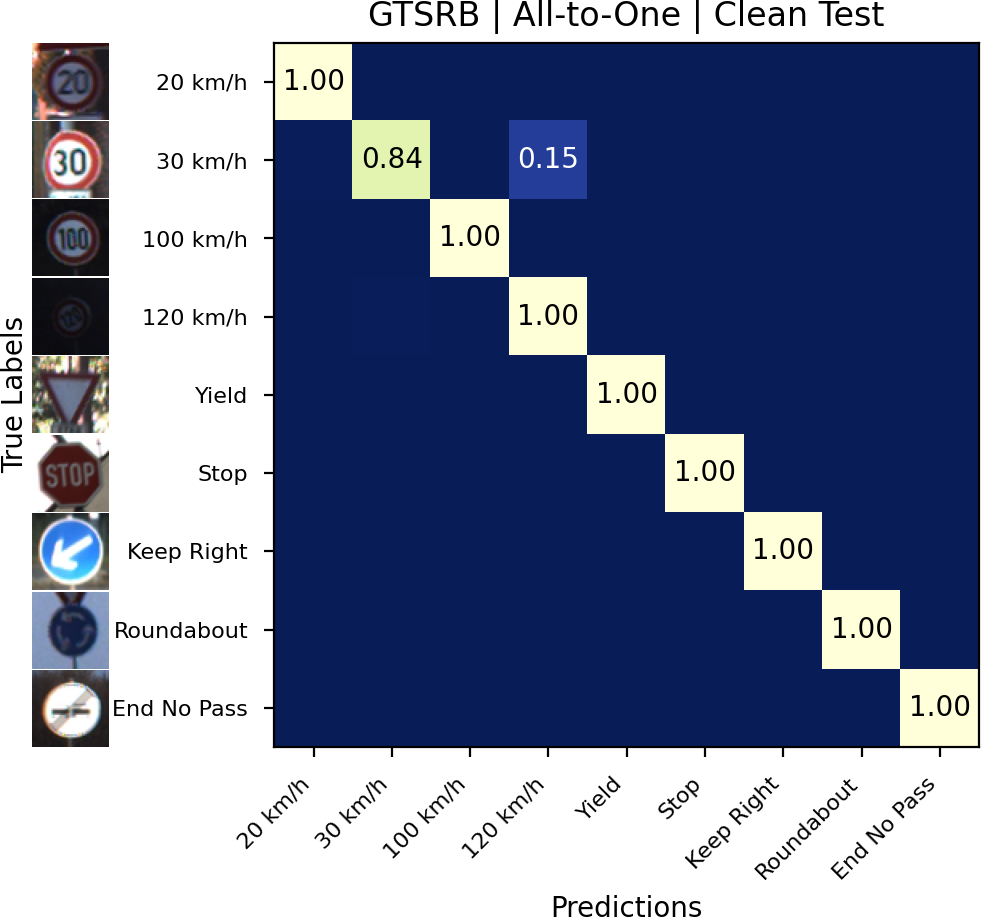}
        \includegraphics[width=0.15\textwidth]{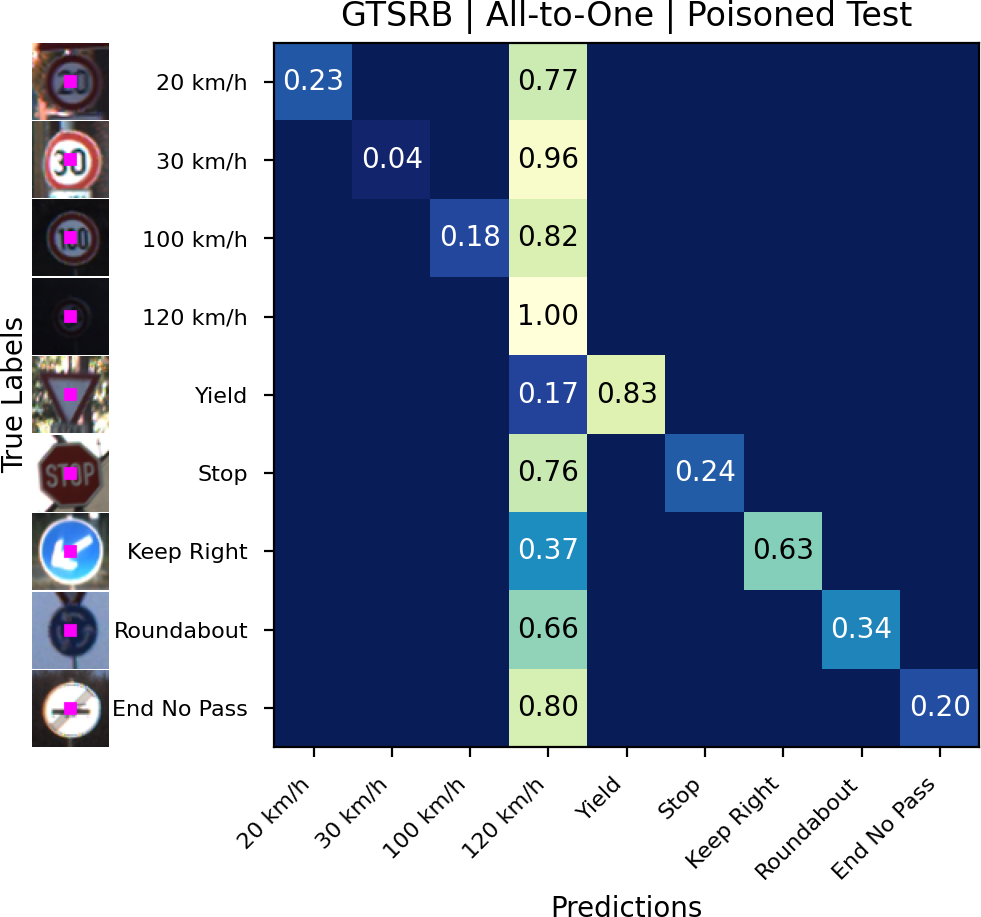}
    \end{minipage}

    \noindent
    \begin{minipage}[c]{0.12\textwidth}
        \centering
        \textbf{CALISA}
    \end{minipage}
    \hfill
    \begin{minipage}[c]{0.85\textwidth}
        \centering
        \includegraphics[width=0.15\textwidth]{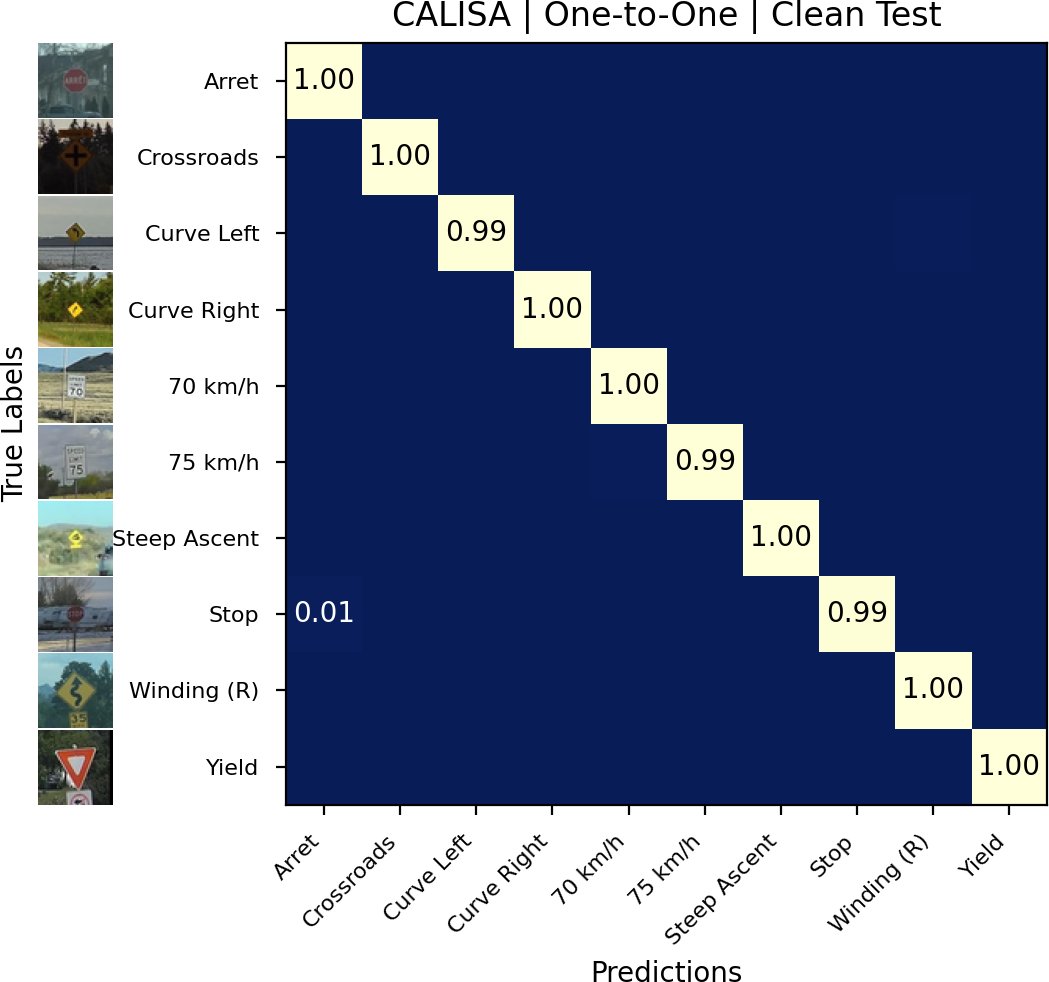}
        \includegraphics[width=0.15\textwidth]{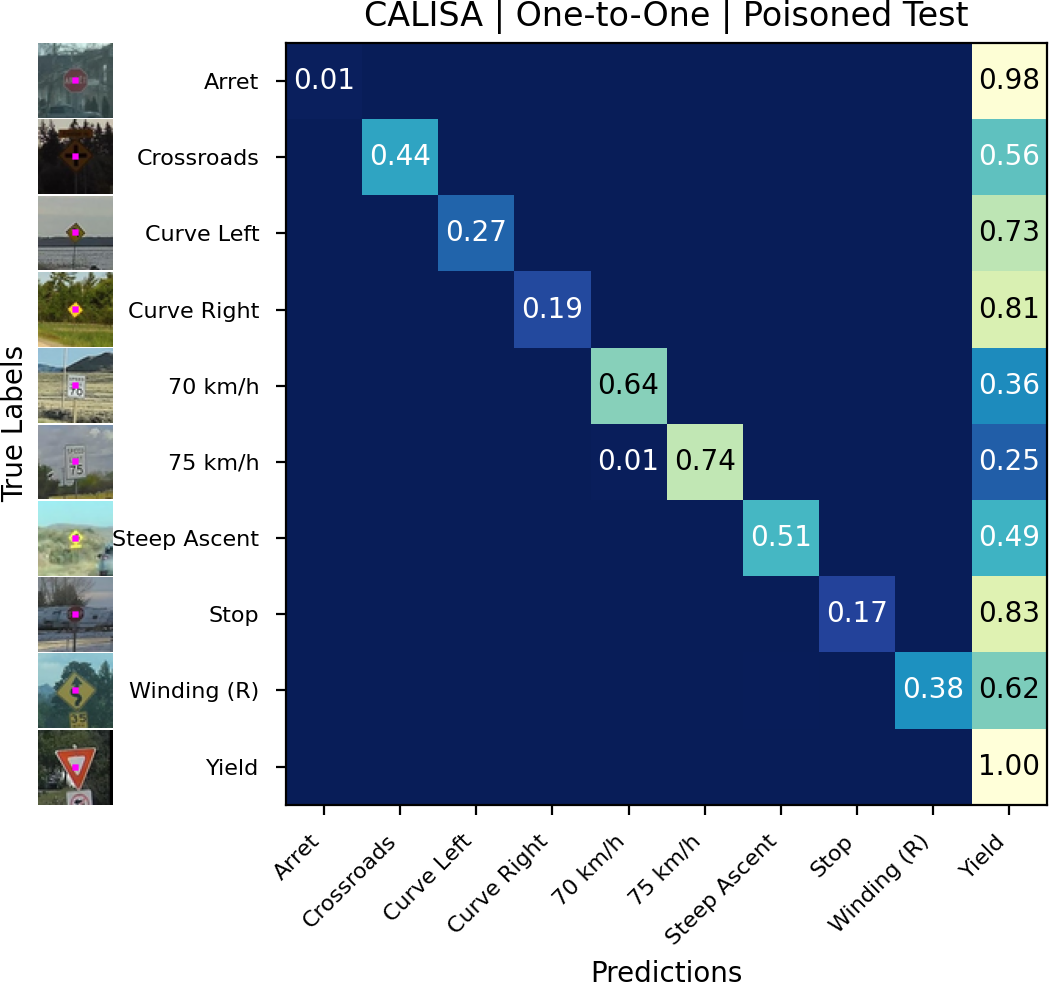}
        \includegraphics[width=0.15\textwidth]{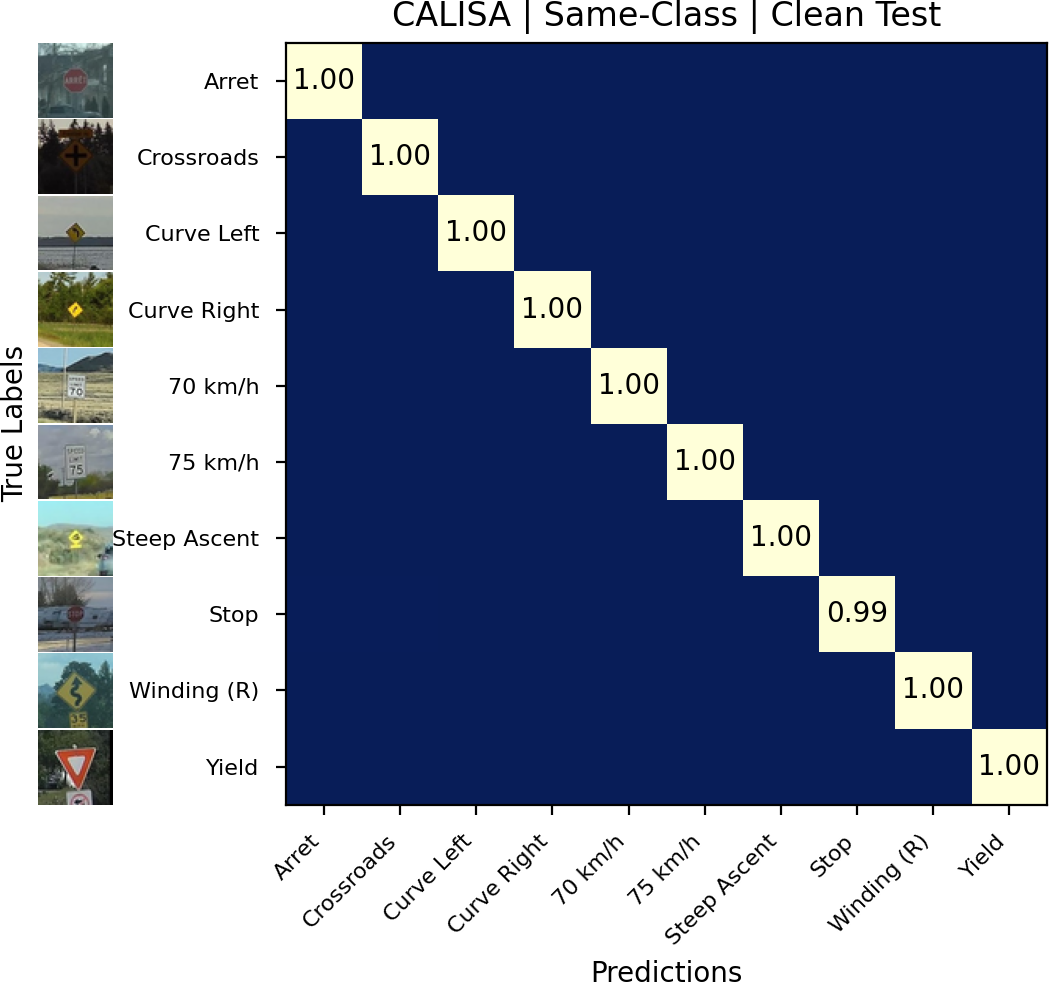}
        \includegraphics[width=0.15\textwidth]{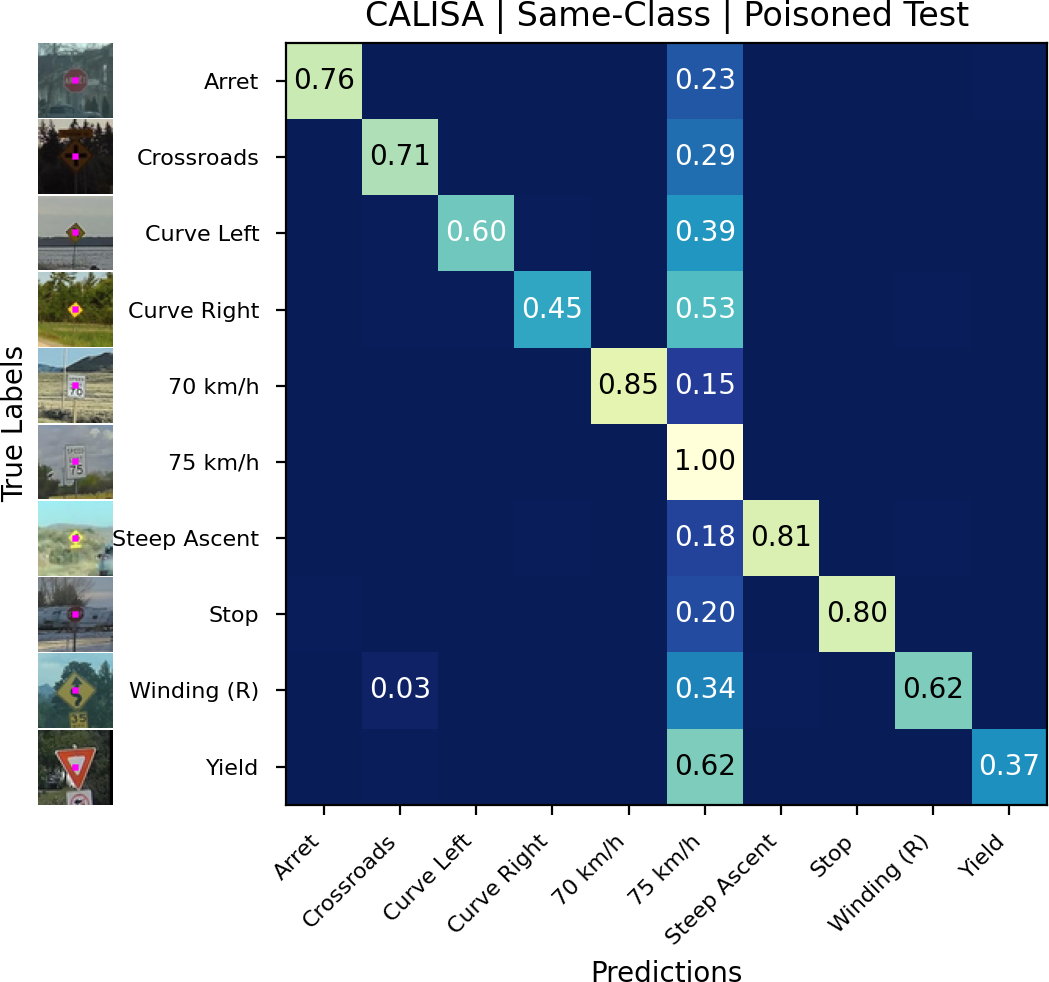}
        \includegraphics[width=0.15\textwidth]{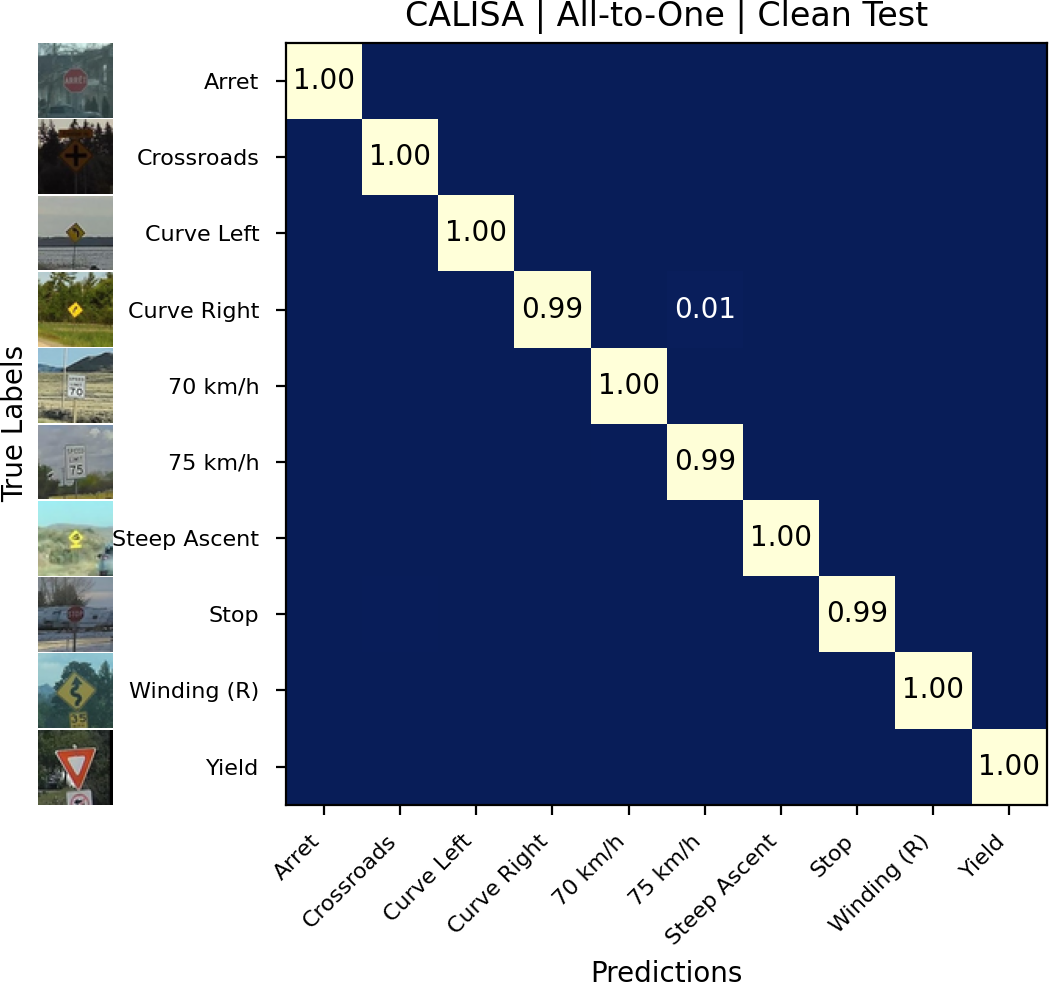}
        \includegraphics[width=0.15\textwidth]{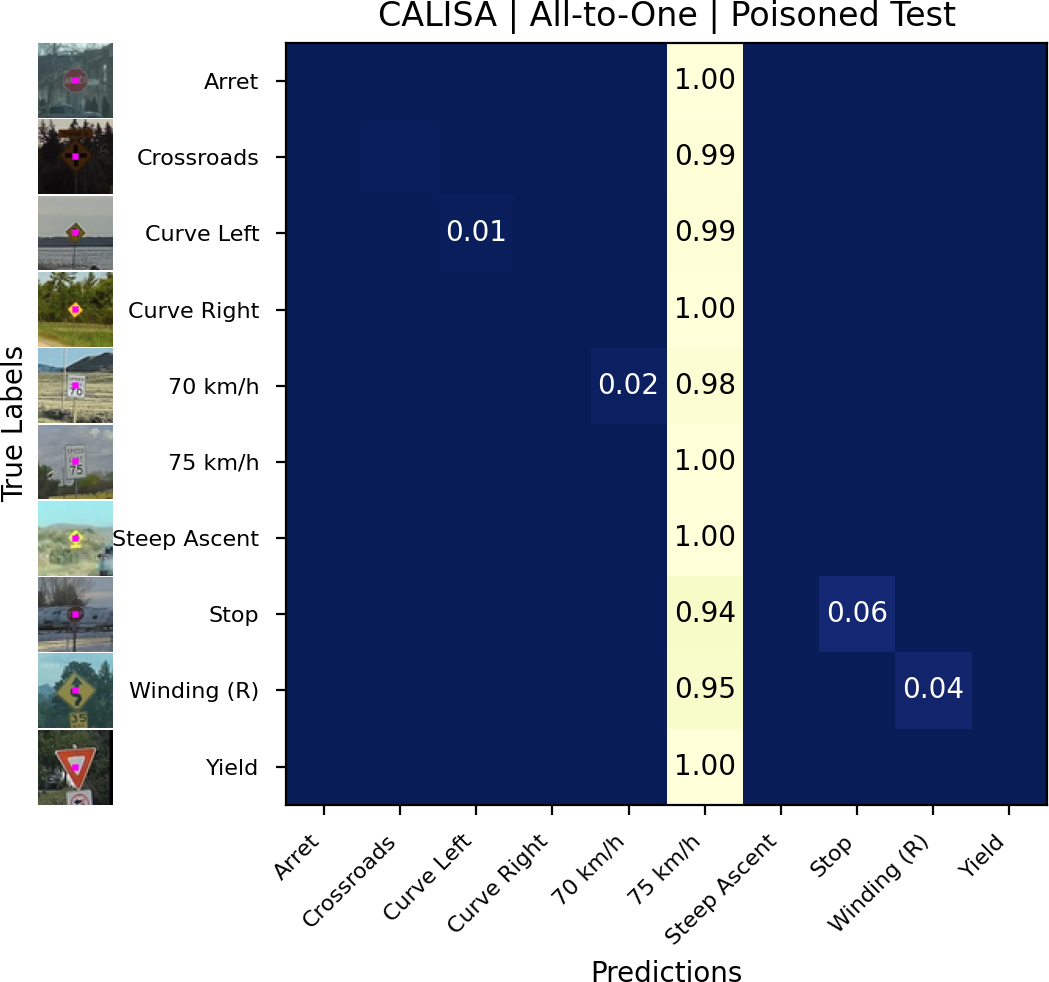}
    \end{minipage}

    \caption{Confusion matrices for the ResNet models trained on DDIM samples.}
    \label{fig:cms_ddim}
\end{figure*}

Table~\ref{tab:attack_fid_ddpm_ddim} provides an overview on the backdoor noise and clean noise backdoor rates for the explored combinations on the victim class versus all other classes (Arb). We also report the change in FID performance compared to a non-backdoored diffusion model. These models exhibit the desired backdoor behaviour only on the victim class (with the exception of one-to-one GTSRB, whose Arb\ssss{nc} is 0.6\%. In all cases, the $\Delta$ impact on the FID score is negligible. Due to space constraints, we include these scores for additional solvers in Appendix~\ref{app:additional-samplers}.

Additionally, Table~\ref{tab:combined_results_side_by_side} reports clean accuracy and attack success rate (ASR) on CIFAR10, GTSRB, and CALISA under different poisoning configurations. We report validation and test performance, with $\Delta$ values measured relative to the train-data baseline on the test split.

Across all datasets, backdoored models maintain competitive performance to both clean synthetic data as well as clean real data, while exhibiting substantially increased ASR. On CIFAR10, poisoning reduces test accuracy by approximately 12\% across attack types, while ASR increases dramatically, reaching over 90\% for one-to-one, same-class, and all-to-one attacks. This demonstrates a successful backdoor despite moderate degradation compared to clean performance. On GTSRB and CALISA, the impact against the target task is even more limited, remaining below 2\% for all attack configurations. In contrast, ASR remains high: all-to-one attacks achieve 98.5\% ASR on GTSRB and 86.9\% on CALISA, while one-to-one attacks reach 80.3\% and 100\% ASR, respectively. Same-class attacks are less effective on both GTSRB and CALISA overall.

Figure~\ref{fig:cms_ddim} shows confusion matrices for CIFAR10, GTSRB, and CALISA under clean and backdoored test conditions for the one-to-one, same-class, and all-to-one poisoning settings using the DDIM sampling method. Due to space constraints, we only provide a subset of the relevant classes for GTSRB and CALISA. On clean test data, the confusion matrices show that benign classification behaviour is largely preserved across all datasets and poisoning settings. In particular, all datasets exhibit strong diagonal dominance, indicating that the poisoning process does not substantially degrade clean classification performance. Limited confusion appears only between visually similar classes (e.g., \textit{cat}/\textit{dog} in CIFAR10) and matches expected clean-training behaviour, consistent with the high clean accuracies reported in Table~\ref{tab:combined_results_side_by_side}.

Under backdoored test conditions, the confusion matrices show that the attack objective is achieved for the same-class and all-to-one scenarios. However, the one-to-one setting does not strongly exhibit the intended goal of maintaining correct class prediction whenever the trigger is applied to arbitrary classes. As a result, we conclude that the current targeted attacks within TEMPO-Diffusion are effective for poisoning downstream models in an all-to-one manner, but are not yet suitable for downstream one-to-one backdoor attacks.

\begin{table}[t]
\centering
\scriptsize
\setlength{\tabcolsep}{7pt}
\renewcommand{\arraystretch}{1.12}
\caption{Recovery-rate and backdoor-rate increase over baseline.}
\label{tab:lambda_recovery_delta}
\begin{tabular}{l c c c c c}
\toprule
Dataset & \shortstack{DDIM\\Clean Rate} & \shortstack{ELIJAH\\$\lambda$} & \shortstack{Recovery\\Rate} & \shortstack{$\Delta$\\Backdoor Rate}& \shortstack{Standard\\Deviation} \\
\midrule
\multirow{4}{*}{\textbf{CIFAR10}}
& \multirow{4}{*}{13.30}
& 0.25 & 0/15  & --     & -- \\
& & 0.50 & 0/15  & --     & -- \\
& & 0.75 & 1/15  & +6.01  & -- \\
& & 1.00 & 6/15 & +32.63 & 22.96 \\
\midrule
\multirow{4}{*}{\textbf{GTSRB}}
& \multirow{4}{*}{13.53}
& 0.25 & 15/15 & +21.05 & 13.90 \\
& & 0.50 & 15/15 & +20.65 & 13.43 \\
& & 0.75 & 15/15 & +16.21 & 13.85 \\
& & 1.00 & 13/15  & +59.81 & 25.56 \\
\midrule
\multirow{4}{*}{\textbf{CALISA}}
& \multirow{4}{*}{10.67}
& 0.25 & 11/15 & +12.12 & 6.08 \\
& & 0.50 & 12/15 & +12.77 & 8.25 \\
& & 0.75 & 13/15 & +21.54 & 15.26 \\
& & 1.00 & 12/15 & +39.01 & 31.38 \\
\bottomrule
\end{tabular}
\end{table}

\subsection{Results: Defence via Trigger Reconstruction}

The main limitation of trigger reconstruction in our threat model is not whether ELIJAH can recover a trigger, but whether that recovery is practical and actionable. In our attack scenario, ELIJAH's trigger inversion is carried out for one class label at a time, and each recovered trigger then requires DDIM sampling to produce 512 images for inspection. As a result, a defender without prior knowledge of the victim class would need to repeat the full procedure across candidate classes and multiple inversion restarts, followed by manual inspection of the generated outputs. This makes ELIJAH computationally expensive against targeted, subtle, multi-output, and input-free backdoors such as TEMPO-Diffusion.

Table~\ref{tab:lambda_recovery_delta} shows that the original ELIJAH setting of $\lambda=0.5$ is not the most effective choice for our poisoned models, but more importantly, recovery rate alone overstates the practical effectiveness of the defence. Although the recovery rate is often high, the corresponding $\Delta$ reflects whether the recovered trigger actually induces backdoor behaviour strongly enough to be immediately recognizable during inspection. On CIFAR10, $\lambda=0.5$ is entirely unsuccessful, with a recovery rate of 0\%, while increasing $\lambda$ to 1.0 raises the recovery rate yields a larger backdoor activation rate. On GTSRB and CALISA, recovery rates are substantially higher across all $\lambda$ values, but many of these recovered triggers still produce relatively modest activation rates. Thus, even when inversion is successful, the defender may still need to inspect many weak or ambiguous generations before identifying a meaningful backdoor pattern. Overall, these results indicate that stronger inversion weighting improves compatibility with our poisoned models, with $\lambda=1.0$ generally yielding the clearest recovered triggers, but the defence remains inefficient and inspection-heavy in practice.

\section{Conclusion}
\label{sec: conclusion}

In this paper, we presented TEMPO-Diffusion, a new class of targeted backdoor attacks on diffusion models capable of generating multiple, sub-image backdoor targets via time-based in-distribution triggers without inference-time access by attackers. Through extensive experiments on CIFAR10, GTSRB, and CALISA datasets, we demonstrated that diffusion models trained under this framework can generate high-fidelity synthetic data that reliably embeds backdoor behaviour in downstream classifiers. We show that diffusion-based synthetic data pipelines can serve as an effective and stealthy vector for downstream poisoning. If diffusion models are used to augment training data in safety-critical pipelines, understanding and mitigating more advanced attack vectors is necessary.

To effectively conduct one-to-one targeted attacks on downstream models, future work should explore backdooring non-target classes during diffusion training while enforcing correct downstream labels. Moreover, beyond single-trigger settings, an important future direction is to extend diffusion backdoors to support multiple concurrent triggers across multiple colour channels and locations.

\bibliographystyle{splncs04}
\bibliography{citations}

\appendix

\section{CALISA Dataset Details}
\label{app:calisa}

Despite the abundance of large vision datasets, datasets for the detection and classification of unique traffic signs remain limited. TT100K~\cite{zhu2016tt100k} and GTSRB~\cite{stallkamp2012gtsrb} are geographically restricted to China and Germany, RTSD~\cite{shakhuro2016rtsd} focuses on Russian signs and is class-imbalanced, MTSD~\cite{ertler2020mapillary} has broad North American coverage but only approximately 2.5\% Canadian data, and LISA~\cite{mogelmose2012lisa} is limited to San Diego and exhibits class imbalance and grayscale bias.

To address these gaps, we introduce CALISA (Canada-Adjusted LISA), a balanced traffic sign dataset spanning Canadian and U.S. signs. CALISA contains 40 classes with 1,000 training and 200 testing images per class, with images sourced directly from Mapillary. The average Canadian proportion is 38.85\%, and 23 classes map directly to LISA labels, covering 5,276 of LISA’s 7,855 instances. Table~\ref{tab:calisa_lisa_canada} details the class mappings and Canadian distribution. To our knowledge, CALISA is the first traffic sign dataset to explicitly emphasize Canadian representation. We constructed CALISA through a three-stage Mapillary v4 Graph API pipeline: crawling the map\_features endpoint over a Canada-wide grid with de-duplication; expanding features into candidate images while removing non-standard perspective images; and decoding traffic-sign detections into bounding boxes. Remaining candidates were then manually verified and filtered.

\begin{table}[t!]
\centering
\caption{CALISA dataset class breakdown.}
\label{tab:calisa_lisa_canada}

\tiny
\setlength{\tabcolsep}{4pt}
\renewcommand{\arraystretch}{1.05}

\begin{minipage}[t]{0.48\linewidth}
\centering
\begin{tabular}{llr}
\toprule
CALISA Name & LISA Name & Canada \% \\
\midrule
arret                   & --                 & 100.00 \\
crossroads              & intersection       & 43.50 \\
curve-left              & curveLeft          & 100.00 \\
curve-right             & curveRight         & 100.00 \\
do-not-pass             & doNotPass          & 0.33 \\
double-curve-left       & --                 & 42.92 \\
double-curve-right      & --                 & 56.67 \\
junction-side-road-left & --                 & 38.75 \\
keep-right              & keepRight          & 100.00 \\
lane-control            & rightLaneMustTurn  & 0.17 \\
merge-right             & addedLane          & 37.92 \\
no-entry                & doNotEnter         & 59.83 \\
no-left-turn            & noLeftTurn         & 3.58 \\
no-u-turn               & --                 & 21.17 \\
one-way-left            & --                 & 5.33 \\
one-way-right           & --                 & 3.08 \\
pedestrian-crossing     & pedestrianCrossing & 68.92 \\
railroad-crossing       & --                 & 0.17 \\
road-narrows-right      & --                 & 100.00 \\
school-zone             & school             & 32.83 \\
\bottomrule
\end{tabular}
\end{minipage}
\hfill
\begin{minipage}[t]{0.48\linewidth}
\centering
\begin{tabular}{llr}
\toprule
CALISA Name & LISA Name & Canada \% \\
\midrule
speedlimit-25       & speedLimit25   & 0.75 \\
speedlimit-30       & speedLimit30   & 29.00 \\
speedlimit-35       & speedLimit35   & 0.33 \\
speedlimit-40       & speedLimit40   & 26.42 \\
speedlimit-45       & speedLimit45   & 1.08 \\
speedlimit-50       & speedLimit50   & 100.00 \\
speedlimit-55       & speedLimit55   & 1.25 \\
speedlimit-60       & --             & 58.42 \\
speedlimit-65       & speedLimit65   & 0.33 \\
speedlimit-70       & --             & 33.33 \\
speedlimit-75       & --             & 0.00 \\
steep-descent       & --             & 13.58 \\
stop                & stop           & 100.00 \\
stop-ahead          & stopAhead      & 54.92 \\
traffic-signals     & --             & 43.75 \\
turn-left           & turnLeft       & 1.50 \\
wild-animals        & --             & 43.17 \\
winding-road-left   & --             & 15.50 \\
winding-road-right  & --             & 15.42 \\
yield               & yield          & 100.00 \\
\bottomrule
\end{tabular}
\end{minipage}
\end{table}

\section{Additional Samplers}
\label{app:additional-samplers}

Due to space constraints, we include the attack and FID scores when using the DPM~\cite{lu2022dpm}, DPM++~\cite{lu2025dpm}, DEIS~\cite{zhang2022deis}, and UniPC~\cite{zhao2023unipc} solvers in Table~\ref{tab:attack_fid_summary_other_solvers}.

\begin{table*}[hb!]
\centering
\caption{Attack performance and FID comparison using $max$ samples under the box-middle setup (trigger size $4\times4$), for DPM, DPM++, DEIS, and UniPC.}
\label{tab:attack_fid_summary_other_solvers}

{\tiny
\setlength{\tabcolsep}{3.5pt}
\renewcommand{\arraystretch}{1.12}

\begin{tabular}{l l c c c c c | c c c c c}
\toprule
& & \multicolumn{5}{c|}{\textbf{DPM}} & \multicolumn{5}{c}{\textbf{DPM++}} \\
\cmidrule(lr){3-7}\cmidrule(lr){8-12}
\textbf{Dataset} & \textbf{Setting}
& \textbf{Vic\textsubscript{nc}} & \textbf{Vic\textsubscript{nb}} & \textbf{Arb\textsubscript{nc}} & \textbf{Arb\textsubscript{nb}} & \textbf{$\Delta$ FID}
& \textbf{Vic\textsubscript{nc}} & \textbf{Vic\textsubscript{nb}} & \textbf{Arb\textsubscript{nc}} & \textbf{Arb\textsubscript{nb}} & \textbf{$\Delta$ FID} \\
\midrule

\multirow{3}{*}{CIFAR10}
& one-to-one & 10.2\% & 96.9\% & 0.0\% & 5.1\% & +1.40 & 16.4\% & 95.3\% & 0.0\% & 5.1\% & +2.32 \\
& same-class & 5.5\% & 96.1\% & 0.0\% & 0.0\% & +0.81 & 4.7\% & 96.9\% & 0.0\% & 0.0\% & +0.45 \\
& all-to-one & 13.3\% & 99.2\% & 0.0\% & 2.7\% & +0.88 & 7.8\% & 96.9\% & 0.0\% & 2.0\% & -0.04 \\
\cmidrule(lr){1-12}

\multirow{3}{*}{GTSRB}
& one-to-one & 14.8\% & 22.7\% & 0.0\% & 4.3\% & +4.61 & 20.3\% & 26.6\% & 0.0\% & 5.9\% & +5.67 \\
& same-class & 12.5\% & 78.9\% & 0.0\% & 0.4\% & +5.87 & 12.5\% & 80.5\% & 0.0\% & 0.8\% & +6.30 \\
& all-to-one & 3.1\% & 88.3\% & 0.0\% & 10.9\% & +6.04 & 7.0\% & 85.2\% & 0.0\% & 6.6\% & +5.27 \\
\cmidrule(lr){1-12}

\multirow{3}{*}{CALISA}
& one-to-one & 4.7\% & 57.8\% & 0.0\% & 5.5\% & +0.51 & 2.3\% & 49.2\% & 0.0\% & 4.4\% & +0.26 \\
& same-class & 11.7\% & 75.8\% & 0.0\% & 0.3\% & +0.24 & 10.2\% & 83.6\% & 0.0\% & 0.0\% & +0.18 \\
& all-to-one & 18.8\% & 77.3\% & 0.0\% & 0.0\% & +0.49 & 18.0\% & 74.2\% & 0.0\% & 0.0\% & +0.57 \\
\midrule

& & \multicolumn{5}{c|}{\textbf{DEIS}} & \multicolumn{5}{c}{\textbf{UNIPC}} \\
\cmidrule(lr){3-7}\cmidrule(lr){8-12}
\textbf{Dataset} & \textbf{Setting}
& \textbf{Vic\textsubscript{nc}} & \textbf{Vic\textsubscript{nb}} & \textbf{Arb\textsubscript{nc}} & \textbf{Arb\textsubscript{nb}} & \textbf{$\Delta$ FID}
& \textbf{Vic\textsubscript{nc}} & \textbf{Vic\textsubscript{nb}} & \textbf{Arb\textsubscript{nc}} & \textbf{Arb\textsubscript{nb}} & \textbf{$\Delta$ FID} \\
\midrule

\multirow{3}{*}{CIFAR10}
& one-to-one & 11.7\% & 95.3\% & 0.0\% & 7.4\% & +1.47 & 18.8\% & 96.1\% & 0.0\% & 6.2\% & +2.08 \\
& same-class & 3.1\% & 96.9\% & 0.0\% & 0.0\% & +0.78 & 6.2\% & 98.4\% & 0.0\% & 0.0\% & +1.31 \\
& all-to-one & 4.7\% & 97.7\% & 0.0\% & 2.3\% & +0.32 & 12.5\% & 96.1\% & 0.0\% & 1.6\% & +0.48 \\
\cmidrule(lr){1-12}

\multirow{3}{*}{GTSRB}
& one-to-one & 13.3\% & 24.2\% & 0.4\% & 6.2\% & +4.65 & 20.3\% & 26.6\% & 0.4\% & 6.6\% & +5.13 \\
& same-class & 12.5\% & 74.2\% & 0.0\% & 0.0\% & +5.75 & 10.9\% & 76.6\% & 0.0\% & 0.0\% & +5.73 \\
& all-to-one & 6.2\% & 86.7\% & 0.0\% & 7.4\% & +5.62 & 6.2\% & 89.1\% & 0.0\% & 14.1\% & +5.95 \\
\cmidrule(lr){1-12}

\multirow{3}{*}{CALISA}
& one-to-one & 2.3\% & 53.9\% & 0.0\% & 4.2\% & +0.20 & 3.1\% & 52.3\% & 0.3\% & 7.6\% & +0.20 \\
& same-class & 10.2\% & 79.7\% & 0.0\% & 0.0\% & +0.04 & 10.2\% & 76.6\% & 0.0\% & 0.0\% & +0.47 \\
& all-to-one & 9.4\% & 75.0\% & 0.0\% & 0.0\% & +0.38 & 17.2\% & 78.9\% & 0.0\% & 0.0\% & +0.35 \\
\bottomrule
\end{tabular}
}
\end{table*}

\end{document}